\documentclass[12pt]{article}
\usepackage{amsmath}
\usepackage{amssymb}
\usepackage{theorem}
\usepackage{euscript}

\newcommand{\PP}{\ensuremath{\mathbb P}}

\newcommand{\ooo}{\ensuremath{\hspace{-0.08cm}\otimes\hspace{-0.08cm}}}

\newcommand{\RR}{\ensuremath{\mathbb R}}
\newcommand{\co}{\ensuremath{|c_{\perp}|}}

{\begin{list}{}{%
\settowidth{\labelwidth}{\textrm{#1~}}%
\setlength{\leftmargin}{\labelwidth+\labelsep}}}%requires macro calc.sty
{\end{list}}
\newtheorem{theorem}{Theorem}[section]
\newtheorem{lemma}[theorem]{Lemma}

\newtheorem{proposition}[theorem]{Proposition}

\theoremstyle{plain}{\theorembodyfont{\rmfamily}%
}

\theoremstyle{plain}{\theorembodyfont{\rmfamily}%
}

\def\blackbox{\leavevmode\vrule height 5pt width 4pt depth 0pt\relax}
\def\endproof{\null\hfill {$\blackbox$}\bigskip}  

\theoremstyle{plain}{\theorembodyfont{\rmfamily}%
\newtheorem{remark}[theorem]{Remark}}
\numberwithin{equation}{section}
\def\blackbox{\leavevmode\vrule height 5pt width 4pt depth 0pt\relax}
\def\endproof{\null\hfill {$\blackbox$}\bigskip}  
\def\eps{\varepsilon}

\begin{document}

\title{The Vlasov model under large magnetic fields in the low-Mach number regime}

\author{P. Degond$^{(1,2)}$, S. A. Hirstoaga$^{(1,2,3)}$, M-H. Vignal$^{(1,2)}$}  
                
\date{May 14th, 2009}
\maketitle

\begin{center}
(1)\, Universit\'e de Toulouse; UPS, INSA, UT1, UTM; Institut de
Math\'ematiques de Toulouse;
F-31062 Toulouse, France 
\end{center}

\begin{center}
(2) CNRS; \, Institut de Math\'ematiques de Toulouse
UMR 5219; F-31062 Toulouse, France. 
\end{center}

\begin{center}
(3) Centre de recherche INRIA Nancy - Grand Est; 
615, rue du Jardin Botanique, F-54602 Villers-L\`es-Nancy, France.
\end{center}

\begin{center}
Emails: pierre.degond@math.univ-toulouse.fr, sever.hirstoaga@math.univ-toulouse.fr, mhvignal@math.univ-toulouse.fr,  
\end{center}

%{\it Key words:}

\begin{abstract}\noindent
This article is concerned with the kinetic modeling, by means of the Vlasov equation, of 
 charged particles under the influence of a strong external electromagnetic
field, i.e. when $\varepsilon^2$,
the dimensionless cyclotron period, tends to zero. 
This leads us to split the velocity variable in the Vlasov equation into fluid and random components. The latter is supposed to have a large magnitude of order $1/\varepsilon$
(which corresponds to the low Mach number regime). 
In the limit $\varepsilon \to 0$, the resulting model is a hybrid model 
which couples a kinetic description of the microscopic random motion of the particles
to a fluid description of the macroscopic behavior of the plasma. The microscopic model
is a first-order partial differential system for the
distribution function, which is averaged over the ultra-fast Larmor gyration and the fast parallel motion along the magnetic field lines. The perpendicular component (with respect to the 
magnetic field lines) of the bulk velocity is governed by the classical relations describing the $E \times B$ and diamagnetic drifts, while its parallel component satisfies an elliptic equation along the magnetic field lines.
\end{abstract}

%%%%%%%%%%%%%%%%%%%%%%%%%%%%%%%%%%%%%%%%%%%%%%%%%%%%%%%%%%%%%%%
%%%%%%%%%%%%%%%%%%%%%%%%%%%%%%%%%%%%%%%%%%%%%%%%%%%%%%%%%%%%%%%
%%%%%%%%%%%%%%%%%%%%%%%%%%%%%%%%%%%%%%%%%%%%%%%%%%%%%%%%%%%%%%%
%%%%%%%%%%%%%%%%%%%%%%%%%%%%%%%%%%%%%%%%%%%%%%%%%%%%%%%%%%%%%%%
%%%%%%%%%%%%%%%%%%%%%%%%%%%%%%%%%%%%%%%%%%%%%%%%%%%%%%%%%%%%%%%

\setcounter{equation}{0} 

\section{Introduction}
In this article, we are interested in the description of the dynamics of
charged particles submitted to a large non uniform magnetic field.
This problem is of great potential interest for describing strongly magnetized plasmas
such as those encountered in Tokamak devices like ITER. The study of plasma confinement
due to a large magnetic field requires to solve the Maxwell equations coupled to the
description of the plasma turbulent transport. This transport can be modelled by using
either a fluid description \cite{Beer_Hammett,deg_san_08,Dorland_Hammett,
Falchetto_Ottaviani_PRL04,Garbet_PhPl01,Hammett_Dorland,Naulin_PhPl03,
Ottaviani_Manfredi_PhPl99,Scott_PhPl05,Xu_PhPl00} or a kinetic description 
\cite{bottino,idomura,kim,Lee,Lin,sugama,tran}. Solving
three dimensional fluid equations is certainly the less expensive way to solve the problem.
However, a fluid description usually overestimates turbulent fluxes especially in a weakly
collisional regime encountered in Tokamaks \cite{dimits,grandgirard}. Indeed
in Tokamaks, the plasma is carried to high temperatures. Thus, since the collision frequency
decreases with increasing temperature, the plasma enters in a nearly collisionless regime.

By contrast, the kinetic model provides an appropriate description of turbulent transport 
in a fairly general context,  but it requires
to solve a six dimensional problem (3D in space and 3D in velocity) which leads to a huge 
computational cost. 
To reduce the cost of numerical simulations, it is classical to derive asymptotic models
with a smaller number of variables than the kinetic description (see~\cite{NBAPD} and
references therein). 

Here, we formally derive a new asymptotic model under both assumptions 
of large magnetic fields and low-Mach numbers. Large magnetic fields 
usually lead to the so-called drift-kinetic limit (see
\cite{ant_lane_80,bri_hahm_07,haz_ware_78,haz_mei_03} for physics references 
and \cite{bostan_08,brenier_00,fre_son_97,fre_son_98,fre_rav_son_01,gol_lsr_99} for
mathematical results). In this regime, due to the large applied magnetic
field, particles are confined along the magnetic field lines and their period of rotation
around these lines (called the cyclotron period) becomes small. However, to our knowledge, 
the consideration of both large magnetic fields and low-Mach numbers is new. As we will see, 
considering low-Mach numbers brings a lot of interesting additional features. 

We consider a simplified plasma model in which we focus on the dynamics of the ions. 
The coupling with the electrons is ignored and the electromagnetic field
is assumed to be given. In future works, coupling the ion dynamics with those of the
electrons and with the electromagnetic field is planned. 
To describe a collisionless ion dynamics (collisions can be neglected 
in Tokamaks in a first instance ; of course, there are situations where 
collisions must be included but we shall discard them in the present work), 
we use the Vlasov equation.
In the large magnetic field regime, the Lorentz force term in the Vlasov equation
is scaled by a large parameter, $1/\varepsilon^2$, where $\varepsilon^2$ stands for
the dimensionless ion cyclotron period, i.e. the rotation period of the ion about a 
magnetic field line (or Larmor rotation). The so called drift-kinetic or gyro-kinetic 
regimes are reached when $\varepsilon$ tends to zero (see \cite{haz_ware_78,Lifshitz}).
We shall not dwell on the distinction between the drift and gyro kinetic regimes, as 
we are aiming at a different situation. 

Indeed, in addition to the magnetic field being large, we assume that 
the ion mean velocity in the plasma is much smaller than the sound
speed. In the Vlasov equation, this assumption implies 
that the fluid bulk velocity is much smaller than the magnitude of the random
motion caused by thermal fluctuations \cite{haz_ware_78}. Thus, we separate the slow scale of the fluid velocity from the
fast scale of the random motion and we express the distribution function as a function of the random component of the velocity (which is also the velocity in the rest frame of the fluid) instead of the velocity in the laboratory frame. In the present work, we focus on the case where the Mach number is of order $\varepsilon$, i.e. it scales like the square root of the dimensionless cyclotron period. This scaling is natural since both quantities scale like the square root of the particle mass, and we can view both the large cyclotron freqency and small Mach numbers as a consequence of the small particle inertia. 
It also turns out that this scaling hypothesis gives rise to a rich structure in the asymptotic regime. 
The scaled
Vlasov equation for the distribution of random velocities must be coupled to the fluid momentum equation which provides an equation for the bulk fluid velocity. Therefore, the unknowns of the Vlasov model in these new variables are the distribution of random velocities and the bulk fluid velocity. 
The goal of this paper is to investigate the limit $\varepsilon\to 0$ of the Vlasov model in this scaling.

When $\varepsilon\to 0$, the limit model consists of two sets of equations, one for the distribution function of random velocities, and one for the bulk fluid velocity. The distribution function of random velocities only depends on space, time and two components of the velocity, corresponding to the parallel component along the magnetic field line and the magnitude of the perpendicular velocity. In other words, the distribution function is independent of the (gyro)-phase of the perpendicular velocity in the plane normal to the magnetic field line. This is a consequence of the ultra-fast cyclotron rotation about the magnetic field lines. It is convenient to express the distribution of random velocities in terms of the parallel velocity and the magnetic moment (or adiabatic invariant), which is proportional to the perpendicular energy divided by the magnitude of the magnetic field. 

Now, the distribution function in these new variables satisfies a first order differential system with a constraint. A Lagrange multiplier allows to express this constraint in the differential system. The constraint expresses that the distribution function is constant along the trajectories of the fast parallel motion along the magnetic field lines. This motion is characterized by the constancy of the magnetic moment and of some kind of pseudo-energy in the parallel direction, which are the two adiabatic invariants of this motion. If a global change of variables from the phase space variables to the adiabatic invariants can be found, it is possible to eliminate the Lagrange multiplier of the constraint and to express the model as a transport equation in the space spanned by these invariants. This transport model describes how the electric and pressure forces as well as spatio-temporal variations of the magnetic field induce a slow evolution of the distribution function function in the space of adiabatic invariants. However, it is not always possible to find such a global change of variables, and, in this situation, the formulation of the problem as a constrained transport equation is the only possible expression of the system. Additionally, in most instances, it will provide a more flexible formulation for numerical discretization. 

The equation for the bulk fluid velocity is split in two equations, one for the perpendicular component to the magnetic field lines, one for the parallel component. The perpendicular component is given through an identity which simply relates this velocity to the $E \times B$-drift and diamagnetic drift velocities. The equation for the parallel component is more unusual. It is an ellptic equation which expresses how, in the zero-Mach number limit, the parallel fluid velocity must adjust in order to guarantee that, at any time, the parallel components of the pressure and electric forces balance along the magnetic field lines. This elliptic equation is highly anisotropic because posed on each magnetic field line. It is obtained by obtained through expressing the constraint of zero parallel force thanks to the moments of the distribution function. 

The derivation of the model roughly follows the following steps: we first proceed with formal expansions of the two unknowns in powers of the parameter $\varepsilon$ (Hillbert expansion) and we keep the first three orders. The expansion of the momentum conservation equation readily leads to an equilibrium constraint expressing that, in the zero-Mach number limit,  the pressure force must balance the electromagnetic force. 
This constraint provides an explicit relation for the perpendicular component of the bulk fluid velocity. The parallel component is only given implicitely through this equilibrium constraint. Finding an explicit equation for it requires some moments of the distribution function. The actual computation of this equation is slightly involved. 

Now, carrying the Hilbert expansion procedure through for the distribution function equation is best done if we change the random velocity variable  into a coordinate system consisting of the parallel velocity, the energy, and the angle of rotation (or gyrophase) around the magnetic field line. Thanks to this coordinate change, we show that the leading order term of the distribution function does not depend on the gyrophase. 

Next, we realize that, at each level of the expansion, we are led to inverting the gyrophase averaging operator \cite{haz_ware_78,haz_mei_03}. We show that the inverse operator can only act on functions satisfying a specific solvability condition, namely that their gyrophase average is zero. We find the asymptotic model in abstract form by imposing this solvability condition successively to the various terms of the expansion, following the classical Hilbert expansion procedure of kinetic theory. Providing explicit expressions of the abstract operators appearing as outcomes of the Hilbert expansion procedures requires somehow tedious computations, most of which will be skipped and given for the reader's convenience in an appendix.

The remainder of the paper is organized as follows. In section~\ref{model}, we present the scaling which expresses the assumptions of strong magnetic field and low Mach number regime.
Then, we present and comment the main result of this article, namely the asymptotic model.
In Section~\ref{sec_drift_kinetic_limit_prelim}, by using Hilbert
expansions we derive the equilibrium constraint for the leading order
fluid velocity and we study the equations concerning the leading order
distribution function. In Section~\ref{sec_gyro_model},
we write the abstract asymptotic model and we provide the main computational steps which lead to the explicit partial differential system
for the limit distribution function.
The conservative form of the model (main result) in terms of the magnetic moment is
obtained in Section~\ref{sec_gyro_interpretation}. Finally, in
section~\ref{explicit_eq_for_u}, we give the explicit formula for the perpendicular
fluid velocity and we obtain the elliptic equation for the parallel part.

%%%%%%%%%%%%%%%%%%%%%%%%%%%%%%%%%%%%%%%%%%%%%%%%%%%%%%%%%%%%%%%
%%%%%%%%%%%%%%%%%%%%%%%%%%%%%%%%%%%%%%%%%%%%%%%%%%%%%%%%%%%%%%%
%%%%%%%%%%%%%%%%%%%%%%%%%%%%%%%%%%%%%%%%%%%%%%%%%%%%%%%%%%%%%%%
%%%%%%%%%%%%%%%%%%%%%%%%%%%%%%%%%%%%%%%%%%%%%%%%%%%%%%%%%%%%%%%
\section{The model, the scaling and the main result}
\label{model}

%%%%%%%%%%%%%%%%%%%%%%%%%%%%%%%%%%%%%%%%%%%%%%%%%%%%%%%%%%%%%%%
%%%%%%%%%%%%%%%%%%%%%%%%%%%%%%%%%%%%%%%%%%%%%%%%%%%%%%%%%%%%%%%
\subsection{The Vlasov equation in a strong magnetic field}
\label{Vlasov}

We are interested in the dynamics of a single species of positively charged ions in the plasma. At this stage of the study, 
the coupling with the electrons is discarded and the electromagnetic field is supposed given.
In future work, the model will be expanded by taking into account the coupling with the electrons and with a self-consistant electromagnetic field. 

We are interested in finding the asymptotic limit of the Vlasov equation describing the dynamics of the ions when
they are submitted to a large external magnetic field and where additionally, the thermal fluctuations of the velocity are large compared with the bulk fluid velocity.

Denoting by $m$ the ion mass and by $q$ the positive charge of the ion,
we start from the Vlasov equation
\begin{equation}\label{V_f}
\partial_tf+v\cdot\nabla_xf+
\frac{q}{m}(E+v\times B)\cdot\nabla_vf=0,
\end{equation}
where $f\equiv f(x,v,t)$ is the distribution function and $x\in \Omega \subset \RR^3_x$,
$v\in\RR^3_v$, and $t\in\RR^+$ are respectively the position, velocity, and
time variables. 
The position $x$ is supposed to belong to an open domain $\Omega \subset \RR^3$. 
In addition, the electric field $E\equiv E(x,t)$ and the
magnetic field $B\equiv B(x,t)$ are assumed to be given.

We supplement this equation with incoming boundary conditions on the boundary $\partial \Omega$ of $\Omega$: 
\begin{eqnarray}
f(x,v,t) = f_B(x,v,t), \quad x \in \partial \Omega, \quad v \cdot \nu(x) <0 , 
\label{VBC}
\end{eqnarray}
where $\nu(x)$ is the outward unit normal to $\partial \Omega$ at $x$ and $f_B$ is supposed given. $f_B$ represents the distribution function of incoming particles in the domain $\Omega$. We also prescribe an initial datum 
\begin{eqnarray}
f(x,v,0) = f_I(x,v), \quad x \in  \Omega,  
\label{VIC}
\end{eqnarray}
where $f_I$ is the distribution function of particles initially present inside the domain $\Omega$.  

Next, we introduce a set of characteristic scales from which an appropriate
scaling of equation \eqref{V_f} will be derived. Let $\bar x$  be a
typical length scale of the problem and let $\overline{v}$ be the ion velocity
scale (typically $\overline{v}$ is the thermal velocity of the ions,
$(2k_{\cal B}\overline{T}/m)^{1/2}$, where $k_{\cal B}$ is the Boltzmann
constant and $\overline{T}$ is the temperature scale). The time scale is
therefore $\overline{t}=\bar x/\overline{v}.$ We denote by $\overline{B}\in\RR^+$
the characteristic magnitude of the applied magnetic field and by
$\overline{E}=\overline{v}\overline{B}$ that of the electric field.
Thus, we define the new variables and given fields by 
$$x'=x/\bar x,\;\;v'=v/\overline{v},\;\;t'=t/\overline{t},\quad E'(x',t')=
E(x,t)/\overline{E},\;\;B'(x',t')=B(x,t)/\overline{B}.$$
Subsequently, letting $\overline{f}$ the distribution function scale, we
introduce the new unknown $f'(x',v',t')=f(x,v,t)/\overline{f}$.

Inserting all these changes into \eqref{V_f} and dropping the primes
for clarity, we obtain the dimensionless
equation
\begin{equation}\label{primes}
\partial_{t}f+v\cdot\nabla_{x}f+\frac{q\overline{B}\overline{t}}{m}(E+
v\times B)\cdot\nabla_{v}f=0.
\end{equation}
When the external magnetic field is assumed to be large, 
the rotation period of the ions about the magnetic field
lines becomes small. Denoting
by $\overline{\omega_c}=\frac{q\overline{B}}{m}$ the characteristic ion
cyclotron frequency, we introduce the
dimensionless cyclotron period 
$$\varepsilon^2=\frac{1}{\overline{t}\,\overline{\omega_c}}.$$
Then, under this scaling, the Vlasov equation \eqref{primes} for
$f=f_\varepsilon$ takes the form:
\begin{equation}\label{V_varepsilon_f}
\partial_t f^\varepsilon+v \cdot \nabla_x f^\varepsilon +
\frac{1}{\varepsilon^2} (E+v\times B) \cdot \nabla_v f^\varepsilon=0,
\end{equation}
with intial and boundaray conditions still given by (\ref{VBC}) and (\ref{VIC}).

%%%%%%%%%%%%%%%%%%%%%%%%%%%%%%%%%%%%%%%%%%%%%%%%%%%%%%%%%%%%%%%

\subsection{Splitting the drift and fluid velocities}
\label{splitting_velocities}

In many instances, the ion mean velocity in the bulk plasma is much smaller than
the sound speed \cite{haz_ware_78}. To take into account this observation, we assume that the fluid ensemble velocity is much smaller that the random component of the particle velocity.
Therefore, besides the magnetic field being large, another key assumption of the present work is that of a low Mach number. Because of the large magnetic field, the random component of the velocity undergoes a fast motion around the magnetic field lines. We will see that this drives the distribution function towards a state which is (at least on the average) isotropic in the plane normal to the magnetic field line. This contributes to a reduction of the dimension of the problem. 

To implement this idea, we need to decompose the velocity into the fluid ensemble velocity and its kinetic part. For this purpose, we define the local density $n^\varepsilon(x,t)$ and fluid velocity $u^\varepsilon(x,t)$ as follows: 
\begin{equation}
n^\varepsilon(x,t) = \int_{\RR^3} f^\varepsilon(x,v,t)\,dv, \quad n^\varepsilon u^\varepsilon(x,t) = \int_{\RR^3} f^\varepsilon(x,v,t) \,v\,dv.
\label{moments}
\end{equation}
The decomposition of the particle velocity is performed through the change of variables 
\begin{equation}
v = u^\varepsilon(x,t) + c/\varepsilon,
\label{chgvar}
\end{equation}
where $c/\varepsilon$ is the random component of the particle velocity. Here the random velocity has a large magnitude $1/\varepsilon$ which is expressed in the choice of the scaling. The distribution function expressed in these new variables is denoted by 
\begin{equation}
f^\varepsilon(x,v,t) = \varepsilon^{3} g^\varepsilon(x,c,t).
\label{f_to_g}
\end{equation}
The scaling factor $\varepsilon^{3}$ in front of $g^\varepsilon$ preserves the density: 
$$\int g^\varepsilon (x,c,t)\,dc=\int f^\varepsilon (x,v,t)\,dv.$$
Because of the definition of the velocity $u^\varepsilon$ according to (\ref{moments}), $g^\varepsilon$ satisfies the constraint: 
\begin{equation}
\int g^\varepsilon (x,c,t) \, c \,dc =0. 
\label{contr}
\end{equation}
Therefore, $g^\varepsilon$ does not carry any information about the evolution of the mean velocity itself. So, the transformed equation from (\ref{V_varepsilon_f}) through the change of variables (\ref{chgvar}) will consist of two parts, an equation for $g^\varepsilon$ on the one hand and the momentum conservation equation which allows to determine $u^\varepsilon$ on the other hand. 

After some easy algebraic manipulations, we find that (\ref{V_varepsilon_f}) is equivalent to the following system: 
\begin{eqnarray}
&&\hspace*{-2cm}\partial_t g^\varepsilon + u^\varepsilon \cdot \nabla_x g^\varepsilon - c \cdot (\nabla_x u^\varepsilon) \nabla_c g^\varepsilon + \frac{1}{\varepsilon} \left( c \cdot \nabla_x g^\varepsilon + \frac{\nabla_x\cdot {\mathbb P}^\varepsilon}{n^\varepsilon} \cdot \nabla_c g^\varepsilon \right) + \nonumber \\
& & \hspace*{8cm} + \frac{1}{\varepsilon^2} (c \times B) \cdot \nabla_c g^\varepsilon = 0,\label{V_varepsilon_g}\\
&&\hspace*{-2cm} \partial_t (n^\varepsilon u^\varepsilon) + \nabla_x \cdot (n^\varepsilon u^\varepsilon \otimes u^\varepsilon)  + \frac{1}{\varepsilon^2} \Bigl( \nabla_x \cdot \PP^\varepsilon - n^\varepsilon (E + u^\varepsilon \times B) \Bigl) = 0,
\label{M_varepsilon_u}
\end{eqnarray}
where $\nabla_x u^\varepsilon$ stands for the Jacobian matrix $(\partial_{x_i}u^\varepsilon_j)_{i,j}$ and $\PP^\varepsilon$ for the pressure tensor,
$$\PP^\varepsilon =\int g^\varepsilon \, c\ooo c\,dc.$$

The model is supplemented with incoming boundary conditions: 
\begin{eqnarray}
g^\varepsilon(x,c,t) = g^\varepsilon_B(x,c,t), \quad x \in \partial \Omega, \quad (u^\varepsilon + \frac{1}{\varepsilon} c ) \cdot \nu(x) <0 , 
\label{VBCg}
\end{eqnarray}
where $g^\varepsilon_B(x,c,t)$ is related to $f_B$ through (\ref{chgvar}), and with initial conditions 
\begin{eqnarray}
g^\varepsilon(x,c,0) = g^\varepsilon_I(x,c), \quad x \in  \Omega,  
\label{VICg}
\end{eqnarray}
where again, $g_I$ is related to $f_I$ through (\ref{chgvar}).

The statement that eqs. (\ref{V_varepsilon_f}), (\ref{moments}) on the one hand, and (\ref{V_varepsilon_g}), (\ref{M_varepsilon_u}) on the other hand, are equivalent, requires some comment. First, (\ref{V_varepsilon_f}), (\ref{moments})  clearly imply (\ref{V_varepsilon_g}), (\ref{M_varepsilon_u}). Indeed, eq. (\ref{M_varepsilon_u}) is simply obtained by multiplying (\ref{V_varepsilon_f}) by $v$ and integrating over $v$. Eq. (\ref{V_varepsilon_g}) is derived by inserting (\ref{chgvar}) and \eqref{f_to_g} into (\ref{V_varepsilon_f}) and using (\ref{M_varepsilon_u}) to eliminate $\partial_t u^\varepsilon$. Conversely, suppose that (\ref{V_varepsilon_g}), (\ref{M_varepsilon_u}) are satisfied. Then, performing the change of variable (\ref{chgvar}) the reverse way leads to (\ref{V_varepsilon_f}). The only thing which remains to be proved is that $u^\varepsilon$ is the average velocity according to definition (\ref{moments}). This is a consequence that the constraint (\ref{contr}) is satisfied by (\ref{V_varepsilon_g}), (\ref{M_varepsilon_u}). To prove this, let us denote by
$I^\varepsilon(x,t)= \int g^\varepsilon(x,c,t) \, c\,dc$. Then, multiplying 
\eqref{V_varepsilon_g} by $c$ and integrating over $c$, we deduce
$$\varepsilon^2 \Big( \partial_t I^\varepsilon +  (u^\varepsilon \cdot \nabla_x) I^\varepsilon + (I^\varepsilon \cdot \nabla_x) u^\varepsilon + (\nabla_x \cdot u^\varepsilon) I^\varepsilon \Big) = I^\varepsilon \times B.$$ 
With the initial condition $I^\varepsilon (x,0)=0$, we obtain
\eqref{contr}. This shows the claimed equivalence.

Our goal is to study the asymptotic model formally obtained by taking the limit
$\varepsilon \to 0$ in \eqref{V_varepsilon_g}, (\ref{M_varepsilon_u}). We notice that taking the first
moment in \eqref{V_varepsilon_g} and closing with $\PP^\varepsilon = p(n^\varepsilon ) \mbox{Id}$, where $n^\varepsilon $ is the density of $g^\varepsilon$, Id is the identity matrix and $p(n)$ is a suitable isentropic pressure law, we find the isentropic  Euler system with Lorentz force, under the scaling used
in \cite{deg_san_08}. This remark guided our choice of the present scaling of the Vlasov system. Indeed, it was shown in \cite{deg_san_08} that this scaling allows to derive a sound drift-fluid model for magnetic plasma confinement.

%%%%%%%%%%%%%%%%%%%%%%%%%%%%%%%%%%%%%%%%%%%%%%%%%%%%%%%%%%%%%%%

\subsection{Main result: the asymptotic model}
\label{sec_main}

We first introduce some notations. We assume that $B$ does not vanish and we define the director of the magnetic field by $b=B/|B|$. For a particle with random component of the velocity $c \in {\mathbb R}^3$, we introduce its parallel component $c_{\parallel}$ with respect to the magnetic field and its magnetic moment $\mu$, given by
$$c_{\parallel}=c\cdot b,\hspace{2cm} \mu=\frac{1}{|B|}\ \frac{|c|^2-c_{\parallel}^2}{2}.$$
We define also define the parallel and perpendicular components of the bulk velocity $u$, by
$$u=u_{\parallel}\,b+u_{\bot},\hspace{1.5cm}u_{\parallel}=u\cdot b, \hspace{1.5cm}u_{\bot}=b\times\left(u\times b\right).$$

The quantity $\mu$ is the magnetic moment of the particle in its rotation motion about the magnetic field, i.e. the magnetic flux through the disk enclosed by the circular trajectory of the particle. 
Indeed, $(|c|^2-c_\parallel^2)/2$ is the kinetic energy of the transverse motion
to $B$ and is proportional to the square of the transverse velocity. But the 
transverse velocity is equal to the Larmor radius $r_L$ times the 
gyrofrequency $\omega_c$ and $r_L^2$ is proportional to the surface $S$ enclosed
by the particle motion while $\omega_c$ is proportional to $|B|$. Thus
$$ \frac{|c|^2-c_\parallel^2}{2} \sim |c_\bot|^2 
\sim r_L^2 \, \omega_c^2 
\sim S \, |B|^2 , $$
where $c_{\bot}=c-c_{\parallel}\,b$ and $\sim$ means proportionality. Thus, $\mu \sim S \, |B|$ which is the magnitude of the flux of $B$ through $S$, as announced. 

For simplicity, we assume that the boundary $\partial \Omega$ of $\Omega$ is a magnetic surface (i.e. at any point on the surface, the magnetic field is tangent to it). As a consequence, a magnetic field line starting inside the interior of $\Omega$ does not intersect the boundary $\partial \Omega$. We also assume that the closure $\bar \Omega$ of $\Omega$ is a compact set. These assumptions are true for instance in most parts of a Tokamak device, since the magnetic field lines are either closed or dense on magnetic surface, depending whether the safety factor is rational or not. We recall that a Tokamak geometry is that of a torus and each magnetic surface at equilibrium is also a torus. The safety factor of a field line is the number of turns around the small circle of the torus which are necessary for one turn around the large one, following the field line. If the safety factor is rational, the field line is closed otherwise the field line is dense on a magnetic surface. The case of magnetic field lines intersecting the boundary can easily be dealt with but will be discarded for simplicity.

The main result of this paper, which will be proven in section~\ref{sec_gyro_model} is the following:

\begin{theorem}\label{main_result}
In the formal limit $\eps\rightarrow 0$, any solution  $(g^\eps, u^\eps)$ of (\ref{V_varepsilon_g}), (\ref{M_varepsilon_u}) converges to $(g,u)$ given as follows: first, there exists a function $G=G(x,\mu,c_{\parallel},t)$ such that
$$
g(x,c,t)=G(x,\mu,c_{\parallel},t).
$$
Defining ${\cal G}=2\,\pi\,|B|\, G$, the functions ${\cal G}$, $u_{\bot}$ and $u_{\parallel}$ are solutions to the following problem: there exists a function  $K=K(x,\mu,c_{\parallel},t)$ such that 
\begin{eqnarray}
& &  \frac{\partial  {\mathcal G}}{\partial t} + {\mathcal S}^\dagger  {\mathcal G} + {\mathcal C}^\dagger  {\mathcal K} = 0 , \label{gyro_cons_1} \\
&&{\mathcal C}^\dagger  {\mathcal G} :=\nabla_x  \cdot ( c_\parallel  {\mathcal G} \, b )  + 
\frac{\partial}{\partial c_\parallel} (( B \cdot \Phi) \, {\mathcal G} )= 0 , \label{gyro_cons_2}\\
& &  u_\bot = \frac{E \times b}{|B|} + \frac{1}{n|B|}(b \times \nabla_x p_\bot)+
\frac{p_\parallel - p_\bot}{n|B|} {\mathbf f}  \,,\label{transv_vel} \\
& &  - 3 (b \cdot \nabla_x) \Bigl( \nabla_x \cdot ( p_\parallel  u_\parallel b) \Bigl)  
+ 2 \, \nabla_x \cdot\Bigl(u_\parallel \nabla_x \cdot (p_\parallel b)\, b\Bigl) + (E \cdot b) \, \nabla_x \cdot (n u_\parallel b) + \nonumber \\
& & \hspace{2cm} + (\nabla_x \cdot b) \nabla_x \cdot \Bigl((-3 p_\parallel + p_\bot) u_\parallel b\Bigl) 
+ p_\bot (\nabla_x \cdot b)^2\, u_{\parallel}  = R_3
\, ,  
\label{eq_u_par} 
\end{eqnarray}
where ${\cal K}=2\,\pi\,|B|\, k$. The right-hand side $R_3$, given by~\eqref{R4}, does not depend on $u_{\parallel}$. Finally, the operator ${\mathcal S}^\dagger$ is defined by 
\begin{eqnarray}
& &  \hspace{-1cm} {\mathcal S}^\dagger   {\mathcal G} :=  \nabla_x \cdot \left[ \left(u + \mu \, \nabla_x \times b - b \times \Phi + (\frac{c_\parallel^2}{|B|} - \mu) \, {\mathbf f} \right)  {\mathcal G} \right] \nonumber \\
& & \hspace{0cm} + \frac{\partial}{\partial c_\parallel} \left[ \left(- (\nabla_x u) : (b \otimes b) + \mu \nabla_x \cdot {\mathbf f} + \Phi \cdot {\mathbf f} \phantom{\frac{1}{|B|}} \hspace{-0.6cm} \right) \, c_\parallel \, {\mathcal G} \right] \nonumber \\
& & \hspace{0cm} + \frac{\partial }{\partial \mu} \left[ \left( - (\frac{\partial}{\partial t} + u \cdot \nabla_x )\ln |B| - \nabla_x \cdot u  + \nabla_x u : (b \otimes b) + \frac{1}{|B|} \nabla_x \cdot (B \times \Phi) \right. \right. \nonumber \\
& & \hspace{8cm} \left. \left. - (  \frac{c_\parallel^2}{|B|} \nabla_x \cdot {\mathbf f} + \Phi \cdot {\mathbf f} ) \right) \mu  \, {\mathcal G} \right]
,
\label{S_dagger} 
\end{eqnarray}
with $\Phi$, ${\mathbf f}$ and ${\mathbb F}$ given by
\begin{equation}
 \Phi = \frac{1}{|B|} ({\mathbb F} - \mu \nabla_x |B|), \hspace{1.5cm}
{\mathbf f} = b  \times (b \cdot \nabla_x) b,  \hspace{1.5cm}
{\mathbb F} = \frac{\nabla_x \cdot {\mathbb P}}{n}, \label{F_agag}
\end{equation}
and with the density $n$ and the pressure tensor ${\mathbb P}$ defined by  
\begin{eqnarray}
& & n = \int {\mathcal G} (\mu ,c_\parallel) \,  d \mu \, dc_\parallel, \quad {\mathbb P} = p_\bot (\mbox{Id} - b \otimes b) + p_\parallel b \otimes b , \label{Pi_P_again} \\
& & p_\bot = \int  {\mathcal G} (\mu ,c_\parallel) \, \mu |B|  \, d \mu \, dc_\parallel , \quad  p_\parallel = \int {\mathcal G} (\mu,c_\parallel) \,  c_\parallel^2 \,   d \mu \, dc_\parallel  ,\label{Pi_pressions_again}
\end{eqnarray}
and the function ${\mathcal K}$ satisfies
\begin{equation}
\int {\mathcal K} \, c_\parallel \, d \mu \, d c_\parallel = 0
 . \label{Kflux=0}
\end{equation}

\end{theorem}

%%%%%%%%%%%%%%%%%%%%%%%%%%%%%%%%%%%%%%%%%%%%%%%%%%%%%%%%%%%%%%%

\subsection{Comments on the asymptotic model}
\label{sub_comments}

\subsubsection{The velocity $u$}
\label{subsub_vel}

Eq.~\eqref{transv_vel} provides an explicit relation for the transverse part of the velocity. In this expression, we recognize the $E \times B$ drift in the first term, and the diamagnetic drift in the second term. The third term is a drift term relating pressure anisotropy and the curvature of magnetic field lines. Indeed, $(b \cdot \nabla_x) b$ is proportional to the curvature of the magnetic field line times the first normal to the curve. The vector $b \times (b \cdot \nabla_x) b$ is proportional to the curvature times the binormal to the curve (in the Frenet frame). 

Eq.~\eqref{eq_u_par} is an elliptic equation for the parallel component $u_\parallel$ of $u$ which is posed along the magnetic field line. Indeed, only operators like  $\varphi \to b \cdot \nabla_x \varphi$ or $\varphi \to \nabla_x \cdot ( b \varphi )$ are applied to $u_\parallel$ in  (\ref{eq_u_par}). In the case of closed magnetic field lines, this problem can be uniquely solved. In the cae of dense field lines on a magnetic surface, it is a conjecture that this problem can be uniquely solved in the space of almost periodic functions. The study of this problem will be developped in future work. 

Both equations follow from the relation 
\begin{eqnarray}
- \frac{\nabla_x \cdot {\mathbb P}}{n} + E + u \times B = 0.
\label{vel_constraint}
\end{eqnarray}
which expresses that, in the zero Mach limit, the total force must vanish. When resolving this equation in terms of $u_\bot$, we immediatly get (\ref{transv_vel}). However, the projection of this equation onto the direction parallel to $b$ gives rise to an implicit constraint on $u_\parallel$: 
\begin{eqnarray}
(- \frac{\nabla_x \cdot {\mathbb P}}{n} + E) \cdot b = 0.
\label{vel_const_par}
\end{eqnarray}
To resolve this constraint into an explicit equation for $u_\parallel$, one must take derivatives of this equation with respect to $t$ and eliminate the time derivatives of $n$, $p_\parallel$ and $p_\bot$ which appear from this operation by using velocity moments of the distribution function equation (\ref{gyro_cons_1}). This computation will be detailed at section (\ref{explicit_eq_for_u}). Note that a similar procedure was used at the level of the drift-fluid limit in \cite{deg_san_08}.

\subsubsection{Trajectories of the fast parallel motion}
\label{subsub_fast}

To understand the constrained transport model (\ref{gyro_cons_1}), (\ref{gyro_cons_2}), it is useful to introduce the negative adjoint operator ${\mathcal C}$ to ${\mathcal C}^\dagger$, given, for a function $F = F(x,\mu, c_\parallel)$ by: 
\begin{eqnarray}
&& {\mathcal C}  F :=   c_\parallel b \cdot \nabla_x F  + ( B \cdot \Phi) \frac{\partial F}{\partial c_\parallel} . \label{Op_C}
\end{eqnarray}
Since $\nabla_x \cdot B = 0 $, a simple computation gives 
\begin{eqnarray}
&& {\mathcal C}^\dagger  {\mathcal G} ={\mathcal C}^\dagger (2 \pi |B| G ) = 2 \pi |B| {\mathcal C} G = |B| {\mathcal C} ( \frac{{\mathcal G}}{|B|})
. \label{Op_C_2}
\end{eqnarray}
Therefore, the constraint (\ref{gyro_cons_2}) can be equivalently written
\begin{eqnarray}
&&  {\mathcal C} G = 0,  
\label{CG=0}
\end{eqnarray}
and expresses that $G = {\mathcal G}/(2 \pi {|B|})$ is constant on the characteristics of (\ref{gyro_cons_2}) or (\ref{Op_C_2}). These characteristics are all the curves defined parametrically by $(x=X(\tau),\mu={\mathcal M}(\tau),c_\parallel= C_\parallel(\tau))$ where $(X(\tau),{\mathcal M}(\tau), C_\parallel(\tau))$ 
satisfies the following ODE: 
\begin{eqnarray}
& & \frac{d X}{d\tau} = C_\parallel \, b,
\label{dXdtau} \\
& & \frac{d C_\parallel}{d\tau} = b \cdot \Big({\mathbb F} - {\mathcal M} \nabla_x |B|\Big),
\label{dCpardtau} \\
& & \frac{d {\mathcal M}}{d\tau} = 0.
\label{dMdtau} 
\end{eqnarray}
Here, $t$ is frozen and $\tau$ is a fictitious time which parametrizes the trajectory. The right-hand sides of (\ref{dXdtau})-(\ref{dMdtau}) are evaluated at 
$(X(\tau),{\mathcal M}(\tau),C_\parallel(\tau))$ and \,$t$. 

The integral curves of the ODE system (\ref{dXdtau})-(\ref{dMdtau}) define the leading order trajectories of the particles in their fast parallel motion along the magnetic field lines. Indeed, their parallel velocity is $C_\parallel$ as seen on (\ref{dXdtau}). Similarly, (\ref{dMdtau}) states that their magnetic moment is invariant in this motion. Finally, (\ref{dCpardtau}) shows that the parallel force is a combination of the projections onto the magnetic field lines of the electric force $E$ on the one hand, and of the mirror force $-\mu\nabla_x|B|$ on the other hand (see \cite[Chapter~4]{haz_mei_03}). This last term takes into account the fact that, if the norm of the magnetic field increases along the particle motion, more energy is converted into the rotation motion about $B$, which decreases the magnitude of the parallel velocity. Symmetrically, if this norm decreases along the particle path, energy is transferred from the rotation to the translation motion and the parallel velocity increases. Now, the constraint (\ref{CG=0}) reflects the fact that the fast parallel motion along the magnetic field line is so fast that it instantaneously relaxes $G$ to a constant along these trajectories. 

More precisely, there are three time scales for a particle moving in a large magnetic field: the fastest time scale (which we will refer to as the 'ultra-fast' time scale) corresponds to the cyclotron or Larmor rotation period about the magnetic field. Then, the second fastest scale (referred to as the fast time scale) is the scale of the parallel motion along the magnetic field line, and which is described by system  (\ref{dXdtau})-(\ref{dMdtau}). Then, there is a slow time scale, which is that of the various drifts across the magnetic field lines, due to spatio-temporal variations of the magnetic field, or to the electric or pressure forces, etc. In this model, we focus on the slow time scale and assume that the ultrafast and fast time scales result in the homogenization (through time averaging) of the distribution function $G$. Therefore, as a consequence of the avearging over the ultra-fast time scale, $G$ is independent of the gyrophase $\alpha$. And because of the fast parallel motion, $G$ is constant along the trajectories (\ref{dXdtau})-(\ref{dMdtau}). 

While expressing this averaging effect directly on the particle trajectories is difficult, the use of the kinetic model directly provides a way to do it by imposing constraints on the distribution function. This easier derivation reflects the fact that, to some extent, the distribution function describes the particle dynamics in a weak (or equivalently statistical) sense. Averaging the trajectories over some fast component is best done by looking at the evolution of an observable of the system which is constant over this fast motion. Saying that $G$ is independent of the gyrophase and is constant along the trajectories of the fast parallel motion is equivalent to saying that there is an equal probability fo find particles with different gyrophases, or at different locations along these trajectories.

\subsubsection{Invariants of the fast parallel motion}
\label{subsub_invar}

The fast parallel motion can be characterized by some invariants. Considering a particular magnetic field line, we introduce its curvilinear abscissa $s$ and we write the function $B \cdot \Phi = b \cdot ({\mathbb F} - \mu \nabla_x |B|)$ locally along this trajectory as the derivative with respect to $s$ of an effective potental ${\mathcal V} (s,\mu)$: 
\begin{eqnarray}
B \cdot \Phi = - \frac{ \partial {\mathcal V} }{ \partial s} (s,\mu). 
\label{cal_v}
\end{eqnarray}
Then, we can write (\ref{dXdtau})-(\ref{dMdtau}) as 
\begin{eqnarray}
& & \frac{d X}{ds} =  b,
\label{dXds} \\
& & C_\parallel \frac{d C_\parallel}{ds} = - \frac{ \partial {\mathcal V} }{ \partial s}(s,{\mathcal M}) ,
\label{dCpards} \\
& & \frac{d {\mathcal M}}{ds} = 0.
\label{dMds} 
\end{eqnarray}
We immediately deduce two integrals of this motion: 
\begin{eqnarray}
{\mathcal M} = \mbox{Constant}, \quad \frac{C_\parallel^2}{2} + {\mathcal V} (s,{\mathcal M}) = \mbox{Constant}. 
\label{integ_par}
\end{eqnarray}
The second relation expresses the conservation of a kind of parallel mechanical energy in the course of the fast parallel motion. 

Therefore, the phase-space trajectories of the fast parallel motion are supported by magnetic field lines for their space dependence, have constant magnetic moment and are such that their parallel velocity is related to the curvilinear abscissa along the magnetic field line and to the magnetic moment according to the second relation of (\ref{integ_par}). The magnetic moment is often referred to as the first adiabatic invariant while the second invariant in (\ref{integ_par}) is related to the second adiabatic invariant. 

Indeed, the second adiabatic invariant is introduced when the fast parallel motion along the magnetic field line has a bounded spatial range. Suppose that we use the second relation of (\ref{integ_par}) to express $C_\parallel$ as a function of the curvilinear abscissa $s$ and that this relation implies that $C_\parallel^2$ can only be non-negative on a bounded interval $[s_1,s_2]$. Then, the particle is bound to oscillate between the two mirror points $s_1$ and $s_2$.  This is because, at $s_1$ and $s_2$, the total energy of the particle, (which is constant along the trajectory) is entirely in the form of perpendicular energy. Then, no energy is available for the parallel motion, and the only possibility for the particle is to bounce back to the regions where the perpendicular energy is lower. This periodic motion is referred to in the physics literature as the bounce motion. By classical mechanical calculations, the period of this motion, or bounce period, which is also called the second adiabatic invariant, is related to the expression appearing in (\ref{integ_par}). However, this expression has a general meaning, while the second adiabatic invariant is only meaningful in the case where the parallel motion is spatially bounded.

\subsubsection{The Lagrange multiplier ${\mathcal K}$}
\label{subsub_K}

Now, ${\mathcal K}$ appearing in (\ref{gyro_cons_1}) is the Lagrange multiplier of the constraint (\ref{gyro_cons_2}). Indeed, let us consider a test function $F (x,\mu, c_\parallel)$ such that 
\begin{eqnarray}
{\mathcal C} F = 0, 
\label{CF=0}
\end{eqnarray}
i.e. such that $F$ is constant along the trajectories of the fast motion. Multiplying (\ref{gyro_cons_1}) by $F$ and integrating with respect to $(x,\mu, c_\parallel)$  leads to 
\begin{eqnarray}
\frac{d}{dt} ({\mathcal G},F) + ({\mathcal S}^\dagger {\mathcal G},F) = 0, 
\label{ddtGF}
\end{eqnarray}
where we have used the notation 
\begin{eqnarray}
({\mathcal G},F) = \int {\mathcal G} \,F \, dx \, d\mu \, dc_\parallel = \int G \,F \, 2 \pi |B| \, dx \, d\mu \, dc_\parallel. 
\label{(G,F)}
\end{eqnarray}
Here, Both $G$ and $F$ are constant along the fast trajectories, and it is equivalent to know $G$ punctually or to know the collections of all $({\mathcal G},F)$ for all test functions $F$ satisfying (\ref{CF=0}). Indeed, the duality formula (\ref{(G,F)}) defines a function $ {\mathcal B} G$: 
\begin{eqnarray}
2 \pi ({\mathcal B}  G,F)= ({\mathcal G},F), \quad \forall F \quad \mbox{ such that } \quad {\mathcal C} F=0. 
\label{calBG}
\end{eqnarray}
Like $G$, the function $ {\mathcal B} G$ is constant on the fast trajectories (note that  ${\mathcal G} = 2 \pi |B| G$ is not~!). Now, by the same considerations, $({\mathcal S}^\dagger {\mathcal G},F)$ determines a function $\overline{{\mathcal S}^\dagger} G$. This operator is unambiguously defined by the relation: 
\begin{eqnarray}
2 \pi (\overline{{\mathcal S}^\dagger} G,F)= ({\mathcal S}^\dagger {\mathcal G},F), \quad \forall F \quad \mbox{ such that } \quad {\mathcal C} F=0. 
\label{barSdagG}
\end{eqnarray}

Is it difficult to write an explicit expression of the operators $ {\mathcal B} G$ and $\overline{{\mathcal S}^\dagger} G$ without a parametrization of the trajectories of the fast motion. As described in section (\ref{subsub_invar}), this requires a parametrization of the magnetic field lines, and a global definition of the two invariants (\ref{integ_par}). Such a global parametrization is not available in general. Section \ref{subsub_invar_var} below examines the particular case where such a global parametrization exists. 

However, definitions (\ref{calBG}) and (\ref{barSdagG}) allow to eliminate the Lagrange multiplier $(\mathcal K)$ and to write (\ref{gyro_cons_1}) according to 
\begin{eqnarray}
\frac{\partial}{\partial t} ({\mathcal B}  G)  + \overline{{\mathcal S}^\dagger} G = 0, 
\label{ddtG}
\end{eqnarray}

The introduction of the Lagrange multiplier is just a convenient way to express the contraint that $G$ is constant along the trajectories of the fast motion, or equivalently, to remove the constraint (\ref{CF=0}) on the test function $F$ for the weak form (\ref{ddtGF}) of (\ref{ddtG}). Indeed, in weak form, system (\ref{gyro_cons_1}), (\ref{gyro_cons_2}) takes the form of a mixed formulation 
\begin{eqnarray}
& & \frac{d}{dt} ({\mathcal G},F) + ({\mathcal S}^\dagger {\mathcal G},F) + ({\mathcal C}^\dagger {\mathcal K},F) = 0, \quad \forall F
\label{ddtGF_mix} \\
& &  ({\mathcal C} G,{\mathcal H}) = 0,  \quad \forall {\mathcal H}
\label{CG_mix}
\end{eqnarray}
where now, the two test functions $F$ and ${\mathcal H}$ are unconstrained. This is the mixed type formulation of the constrained problem (\ref{ddtGF}), (\ref{CF=0}). We note that ${\mathcal K}$ is defined up to a solution of ${\mathcal C}^\dagger {\mathcal K} = 0$. It does not seem obvious to single out a more pertinent choice of ${\mathcal K}$  among these possible solutions.
However, this multiplicity does not affect the problem since whatever choice will lead to the same solution ${\mathcal G}$.

\subsubsection{Expression in the global coordinate system spanned by the invariants (if it exists)}
\label{subsub_invar_var}

To provide more explicit expressions of the operators  $ {\mathcal B} G$ and $\overline{{\mathcal S}^\dagger} G$, we assume that there exists a global change of variables 
\begin{eqnarray}
x \to (y(x),s(x)), 
\label{ch_var_glob_inv}
\end{eqnarray}
such that $y \in {\mathbb R}^2$. A magnetic field line is defined by $y(x) = $ Constant,  and $s(x)$ is the curvilinear abscissa along the magnetic field line. We denote by $(\mu, \mathcal{W})$, the two invariants defined by (\ref{integ_par}) and we also assume that $\mathcal{W}$ can be defined globally as a function $\mathcal{W}(x,\mu,c_\parallel)$. When the vector-valued function $\Psi(x,\mu,c_\parallel)$ defined by 
\begin{eqnarray}
\Psi(x,\mu,c_\parallel)  =  (y(x), \mu, {\mathcal W}(x,\mu,c_\parallel)) , 
\label{def_Psi}
\end{eqnarray}
takes a fixed value $(y_0, \mu_0, {\mathcal W}_0)$, the point $(x,\mu,c_\parallel)$ in phase-space spans a given trajectory of the fast motion denoted by $C_{(y_0, \mu_0, {\mathcal W}_0)}$. We denote by $dl_{(y_0, \mu_0, {\mathcal W}_0)}$ the length element on this curve. 

Now, we obviously can write $G$ as a function of $(y, \mu, {\mathcal W})$: 
\begin{eqnarray}
G(x,\mu,c_\parallel,t) = \bar G (y(x), \mu, {\mathcal W}(x,\mu,c_\parallel)).  
\label{barG}
\end{eqnarray}
As a particular test function, we choose 
\begin{eqnarray}
F(x,\mu,c_\parallel,t) = \delta (\Psi(x,\mu,c_\parallel) - (y_0, \mu_0, {\mathcal W}_0)).  
\label{F=delta}
\end{eqnarray}
By the coarea formula \cite{Federer}, the expression involving the delta distribution means that, for any smooth (for instance, continuous) test function $\phi$, we have 
\begin{eqnarray}
\int \phi \, \delta (\Psi(x,\mu,c_\parallel) - (y_0, \mu_0, {\mathcal W}_0)) \, dx \, d\mu \, dc_\parallel = \int_{C_{(y_0, \mu_0, {\mathcal W}_0)}} \phi \frac{dl_{(y_0, \mu_0, {\mathcal W}_0)}}{\Delta} ,  
\label{delta}
\end{eqnarray}
where $\Delta = \sqrt{ \det ( d\Psi \, d\Psi^*)}$, $d\Psi$ is the derivative of the above defined function $\Psi$, $d\Psi^*$ is the matrix transpose of $\Psi$ and the expression $d\Psi \, d\Psi^*$ means a matrix product. We note that for any function $G$ like (\ref{barG}), the following identity holds true: 
\begin{eqnarray}
\int G \, \delta (\Psi(x,\mu,c_\parallel) - (y_0, \mu_0, {\mathcal W}_0)) \, dx \, d\mu \, dc_\parallel = \Lambda (y_0, \mu_0, {\mathcal W}_0) \, \bar G(y_0, \mu_0, {\mathcal W}_0) 
\label{deltaG}
\end{eqnarray}
where 
\begin{eqnarray}
\Lambda (y_0, \mu_0, {\mathcal W}_0) = \int \, \delta (\Psi(x,\mu,c_\parallel) - (y_0, \mu_0, {\mathcal W}_0)) \, dx \, d\mu \, dc_\parallel.  
\label{Lambda}
\end{eqnarray}
is the coarea (colength) of $C_{(y_0, \mu_0, {\mathcal W}_0)}$. 
The function $F$ given by (\ref{F=delta}) obviously satisfies the constraint (\ref{CF=0}), and therefore, (\ref{barSdagG}) leads to the following expression of 
$\overline{{\mathcal S}^\dagger} G $:
\begin{eqnarray}
(\Lambda \, \overline{{\mathcal S}^\dagger} G) (y_0, \mu_0, {\mathcal W}_0) = \int {\mathcal S}^\dagger {\mathcal G} \, \, 
\delta (\Psi(x,\mu,c_\parallel) - (y_0, \mu_0, {\mathcal W}_0))  \, dx \, d\mu \, dc_\parallel
\label{SdagG_inv}
\end{eqnarray}

Now, let us define by $S_x \in {\mathbb R}^3$, $S_\mu \in {\mathbb R}$, $S_\parallel \in {\mathbb R}$, the fields such that 
\begin{eqnarray}
{\mathcal S}^\dagger {\mathcal G}  = \nabla_x \cdot ( S_x {\mathcal G}) + \frac{\partial}{\partial \mu} ( S_\mu {\mathcal G}) +  \frac{\partial}{\partial c_\parallel} ( S_\parallel {\mathcal G}). 
\label{field_S}
\end{eqnarray}
We write 
$$ {\mathcal S}^\dagger {\mathcal G}  = \nabla_{(x,\mu,c_\parallel)} \cdot (S {\mathcal G}) , $$
with $S=(S_x,S_\mu,s_\parallel)$. 
Then, bu using the Green formula and the chain rule: 
\begin{eqnarray}
& & (\Lambda \, \overline{{\mathcal S}^\dagger} G) (y_0, \mu_0, {\mathcal W}_0) = \int \nabla_{(x,\mu,c_\parallel)} \cdot (S {\mathcal G})
\delta (\Psi(x,\mu,c_\parallel) - (y_0, \mu_0, {\mathcal W}_0))  \, dx \, d\mu \, dc_\parallel \nonumber \\
&=& - \int     {\mathcal G} \, S \cdot \nabla_{(x,\mu,c_\parallel)} \left[
\delta (\Psi(x,\mu,c_\parallel) - (y_0, \mu_0, {\mathcal W}_0)) \right]  \, dx \, d\mu \, dc_\parallel \nonumber \\
&=& - \int     {\mathcal G}   \nabla_{(y,\mu,{\mathcal W})} \left[ \delta ((y, \mu, {\mathcal W}) - (y_0, \mu_0, {\mathcal W}_0)) \right]|_{(y, \mu, {\mathcal W}) = \Psi(x,\mu,c_\parallel)} ( \nabla_{(x,\mu,c_\parallel)} \Psi \cdot S ) \, dx \, d\mu \, dc_\parallel \nonumber \\
&=&  \nabla_{(y_0,\mu_0,{\mathcal W}_0)} \cdot \left( \int     {\mathcal G}     (\nabla_{(x,\mu,c_\parallel)} \Psi \cdot S) \delta (\Psi(x,\mu,c_\parallel) - (y_0, \mu_0, {\mathcal W}_0)) \, dx \, d\mu \, dc_\parallel \right) 
\label{SdagG_inv_2}
\end{eqnarray}
For the last identity, we have used that, 
$$\nabla_{(y,\mu,{\mathcal W})} \left[ \delta ((y, \mu, {\mathcal W}) - (y_0, \mu_0, {\mathcal W}_0)) \right] = - \nabla_{(y_0,\mu_0,{\mathcal W}_0)} \left[ \delta ((y, \mu, {\mathcal W}) - (y_0, \mu_0, {\mathcal W}_0)) \right]. $$

Finally, we can define the vector fields 
$\Sigma_y \in {\mathbb R}^2$, $\Sigma_\mu \in {\mathbb R}$, $\Sigma_{{\mathcal W}} \in {\mathbb R}$ and $\Sigma = (\Sigma_y , \Sigma_\mu, \Sigma_{{\mathcal W}})$, such that 
\begin{eqnarray}
& & \hspace{-1cm} \Sigma_y (y_0, \mu_0, {\mathcal W}_0) =  \int     2 \pi |B| (\nabla_{x} y \cdot S_x) \delta (\Psi(x,\mu,c_\parallel) - (y_0, \mu_0, {\mathcal W}_0)) \, dx \, d\mu \, dc_\parallel ,
\label{Sig_y} \\
& & \hspace{-1cm} \Sigma_\mu (y_0, \mu_0, {\mathcal W}_0) =  \int     2 \pi |B|  S_\mu \delta (\Psi(x,\mu,c_\parallel) - (y_0, \mu_0, {\mathcal W}_0)) \, dx \, d\mu \, dc_\parallel ,
\label{Sig_mu} \\
& & \hspace{-1cm}  \Sigma_{{\mathcal W}} (y_0, \mu_0, {\mathcal W}_0) =  \int     2 \pi |B| (\nabla_{(x,\mu,c_\parallel)} {\mathcal W} \cdot S) \delta (\Psi(x,\mu,c_\parallel) - (y_0, \mu_0, {\mathcal W}_0)) \, dx \, d\mu \, dc_\parallel ,
\label{Sig_W} 
\end{eqnarray}
and write: 
\begin{eqnarray}
\overline{{\mathcal S}^\dagger} G &=& \Lambda^{-1} \nabla_{(y, \mu, {\mathcal W})} \cdot (\Sigma \bar G) . 
\label{SdagG_inv_3}
\end{eqnarray}
Similar considerations lead to the fact that 
\begin{eqnarray}
{\mathcal B} (x,\mu,c_\parallel,t) = \bar {\mathcal B}  (y(x), \mu, {\mathcal W}(x,\mu,c_\parallel)),  
\label{calB=barcalB}
\end{eqnarray}
with 
\begin{eqnarray}
& & \hspace{-1cm} (\Lambda \, \bar {\mathcal B} ) (y_0, \mu_0, {\mathcal W}_0) =  \int     2 \pi |B| \,  \delta (\Psi(x,\mu,c_\parallel) - (y_0, \mu_0, {\mathcal W}_0)) \, dx \, d\mu \, dc_\parallel .
\label{barcalB}
\end{eqnarray}

Finally, system (\ref{gyro_cons_1}), (\ref{gyro_cons_1}) is equivalent to the following system for $\bar G(y,\mu,{\mathcal W}, t)$: 
\begin{eqnarray}
& & \frac{\partial}{\partial t} ({\bar {\mathcal B} \bar G}) + \Lambda^{-1} \nabla_{y,\mu,{\mathcal W}} (\Sigma \bar G) = 0. 
\label{gyro_var_invar} 
\end{eqnarray}
This system is a classical first order hyperbolic equation for $\bar G$.

Note that the existence of a global change of variables like (\ref{def_Psi}) is not always guaranteed. Of course, a local one always exist (using the rectification theorem for vector fields for instance), but it may not be globally defined. Even if it is the case, the practical determination of this change of variables may be a challenging problem, which makes the use of these coordinates very often impossible. If the magnetic field evolves in time, this change of coordinates must be updated at any new time step, whch increases the computational difficulty of the problem even further. 

Another comment is about the finiteness of the integrals (\ref{Lambda}), (\ref{Sig_y}) to (\ref{Sig_W}) and (\ref{barcalB}). If the magnetic field lines are closed, these integrals are on a compact set, and they are finite. If the magnetic field lines are not closed, but are dense on a magnetic surface, then, by imposing smoothness conditions on the test functions $F$, we deduce that the integrals are not defined on a particular magnetic field line, but rather on the whole magnetic surface.Then, we must change the definitions of the coefficients of the system accordingly. But with this change, the integrals become surface integrals on a compact manifold again (since we assumed from the very beginning that the closure $\bar \Omega$ of the domain $\Omega$ was compact), and are finite as well. Of course, this requires to shift from one definition of the coefficients to another one and is very difficult to realize in practice. The weak mixed formulation is more suitable because it allows to automatically shift from line integrals to surface integrals when passing from closed magnetic field lines to dens ones.

\subsubsection{Initial and boundary conditions}
\label{subsub_ibc}

Let us first make some comments on the geometric assumptions of the boundary $\partial \Omega$, namely that $\partial \Omega$ is magnetic surface. It implies that no magnetic field line originating from the interior of $\Omega$ intersects $\partial \Omega$. If this were the case, then constaint (\ref{CG=0}) would impose that the value of $G$ inside the domain along this field line would be specified by the boundary condition at the point where this magnetic field line interesects the boundary. This would result in the fact that the limit model would actually be posed on a subdomain $\Omega' \subset \Omega$ consisting of the union of all magnetic field lines which do not intersect the boundary. In order to avoid any further complicated geometrical discussion, we wilil discard this situation and assume that $\partial \Omega$ is a magnetic field surface.   

The model (\ref{gyro_cons_1}), (\ref{gyro_cons_2}) has to be supplemented with suitable boundary conditions on $\partial \Omega$. The case where there exist global invariants $(y,\mu, {\mathcal W})$ to the trajectories of the fast motion will guide us in determining what are the proper boundary conditions. Indeed, in this case, the model reduces to the first order differential system (\ref{gyro_var_invar}) in the invariant space $(y,\mu, {\mathcal W})$. 

We first specify some notations. The domain $\Omega \times {\mathbb R}_+ \times {\mathbb R}$ where the variable $(x,\mu,c_\parallel)$ takes its values transforms into a domain ${\mathcal O}$ through the transformation to the $(y,\mu, {\mathcal W})$ variables (\ref{def_Psi}). Its boundary is denoted by $\partial {\mathcal O}$. The boundary conditions must be prescribed on the part of $\partial {\mathcal O}$ where the vector field $\Sigma$ is incoming, i.e. $\Sigma \cdot \lambda < 0$  where $\lambda$ denotes the outward unit normal to $\partial {\mathcal O}$ at the considered point on the boundary. 

Let us first assume that $g_B^\varepsilon$ given by (\ref{VBCg}) is such that 
\begin{eqnarray}
& & g_B^\varepsilon \to G_B (x,\mu,c_\parallel,t), \quad x \in \partial \Omega, \quad c \cdot \nu(x) <0 , 
\label{VBCglim} 
\end{eqnarray}
as $\varepsilon \to 0$. We also assume that $G_B$ satisfies ${\mathcal C} G_B = 0$, i.e. is constant alont the phase-space trajectories of the fast motion. This assumption is consistant since, by the geometric assumption on $\partial \Omega$, magnetic field lines starting from a point on $\partial \Omega$ will remain on $\partial \Omega$. 

Then, we note the following identity: 
\begin{eqnarray}
& & \hspace{-1cm} ( \nabla_{(y,\mu,{\mathcal W})} \cdot (\Sigma \bar G) , \bar F)_{\mathcal O} + ( \bar G, \Sigma \cdot \nabla_{(y,\mu,{\mathcal W})} \bar F)_{\mathcal O}
= \int_{\partial {\mathcal O}} \bar G \bar F (\Sigma \cdot \lambda) \, d  \Sigma (y,\mu,{\mathcal W}) , 
\label{dualy} 
\end{eqnarray}
where $(\bar F, \bar G)_{\mathcal O}$ stands for the integral of $\bar F \bar G$ over  ${\mathcal O}$ and $d  \Sigma (y,\mu,{\mathcal W})$ is the superficial measure on $\partial {\mathcal O}$. Similarly, we have 
\begin{eqnarray}
& & \hspace{-1cm} ( \nabla_{(x,\mu,c_\parallel)} \cdot (S {\mathcal G}) ,  F) + ( {\mathcal G}, S \cdot \nabla_{(x,\mu,c_\parallel)} F)
= \int_{\partial \Omega \times {\mathbb R}_+ \times {\mathbb R} } {\mathcal G}  F (S_x \cdot \nu) \, dS(x) \, d \mu \, dc_\parallel , 
\label{dualx} 
\end{eqnarray}
where $dS(x)$ is the superficial measure on $\partial \Omega$ and we recall the definition (\ref{(G,F)}). By the calculation developed in section \ref{subsub_invar_var}, the first and second terms of the left-hand side of (\ref{dualy}) are respectively equal to the first and second terms of the left-hand side of (\ref{dualx}), we deduce that the right-hand sides of these two formulas are equal. Splitting the boundary $\Gamma = \partial \Omega \times {\mathbb R}_+ \times {\mathbb R} $ into $\Gamma_+$, $\Gamma_-$, and $\Gamma_0$, according to whether $(S_x \cdot \nu)$ is $>0$, $<0$ or $=0$, and splitting $\partial {\mathcal O}$ into $\partial {\mathcal O}_+$, $\partial {\mathcal O}_-$, $\partial {\mathcal O}_0$ in a similar way according to the sign of $(\Sigma \cdot \lambda)$, we get 
\begin{eqnarray}
& & \hspace{-1cm} \int_{\partial {\mathcal O}_+} \bar G \bar F (\Sigma \cdot \lambda) \, d  \Sigma (y,\mu,{\mathcal W})  - \int_{\partial {\mathcal O}_-} \bar G \bar F |(\Sigma \cdot \lambda)| \, d  \Sigma (y,\mu,{\mathcal W}) = \nonumber \\
& & \hspace{1cm} 
= \int_{\Gamma_+} {\mathcal G}  F (S_x \cdot \nu) \, dS(x) \, d \mu \, dc_\parallel - \int_{\Gamma_-} {\mathcal G}  F |(S_x \cdot \nu)| \, dS(x) \, d \mu \, dc_\parallel
.
\label{BCsplit} 
\end{eqnarray}

We now make a very strong assumption, without which it is difficult to prescribe sound boundary conditions for this model. We assume that the sign of $(S_x \cdot \nu)$ is constant along the trajectories of the fast motion. Then, it is consistant to assume that the test function $F$ satisfies the constraint (\ref{CF=0}), vanishes identically on $\Gamma_+$ but does not vanish and is actually arbitrary on $\Gamma_-$. For such a test function, we deduce that 
\begin{eqnarray}
& & \hspace{-1cm}  \int_{\partial {\mathcal O}_-} \bar G \bar F |(\Sigma \cdot \lambda)| \, d  \Sigma (y,\mu,{\mathcal W}) =  \int_{\Gamma_-} {\mathcal G}  F |(S_x \cdot \nu)| \, dS(x) \, d \mu \, dc_\parallel
.
\label{BCincom} 
\end{eqnarray}
Then, prescribing $\bar G$ on $\partial \Omega_-$ is equivalent to prescribing the value of the integral at the left-hand side of (\ref{BCincom}) and, thanks to this relation, to prescribing the integral at the right-hand side of (\ref{BCincom}).

Therefore, we prescribe the following boundary condition for the solution ${\mathcal G} $ of (\ref{gyro_cons_1}), (\ref{gyro_cons_2}):
\begin{eqnarray}
& & \hspace{-1cm}  \int_{\Gamma_-} {\mathcal G}  F |(S_x \cdot \nu)| \, dS(x) \, d \mu \, dc_\parallel = \int_{\Gamma_-} 2 \pi |B| G_B   F |(S_x \cdot \nu)| \, dS(x) \, d \mu \, dc_\parallel
.
\label{BCincomGB} 
\end{eqnarray}
for any test function $F$ which satisfies the constraint (\ref{CF=0}) and vanishes identically on $\Gamma_+$, and where the datum $G_B$  is given by (\ref{VBCglim}).

If it is possible to use global invariants as coordinates $(y,\mu,{\mathcal W})$, we can use (\ref{BCincom}) with a test function $F$ of the form (\ref{F=delta}) with $(y_0,\mu_0,{\mathcal W}_0) \in \partial {\mathcal O}_-$. This leads to 
\begin{eqnarray}
& & \hspace{-1cm}   (\bar G  |(\Sigma \cdot \lambda)| \, d  \Sigma) (y_0,\mu_0,{\mathcal W}_0)  = \nonumber \\
& & \hspace{1cm}  \, = \int_{\Gamma_-} 2 \pi |B| G_B |(S_x \cdot \nu)| \, \delta (\Psi(x,\mu,c_\parallel) - (y_0, \mu_0, {\mathcal W}_0)) \, dS(x) \, d \mu \, dc_\parallel ,
,
\label{BCincomGBinvar} 
\end{eqnarray}
which gives an explicit prescription for $\bar G$ on the incoming boundary $\partial {\mathcal O}_-$. 

The prescription (\ref{BCincomGB}), or in the cases where it is possible to use global invariants as coordinates, (\ref{BCincomGBinvar}) are the prescribed boundary conditions for the model (\ref{gyro_cons_1}), (\ref{gyro_cons_2}). 

The initial conditions are easier. We suppose that the initial conditions (\ref{VICg}) are such that there exists $G_I(x,\mu,c_\parallel)$, satisfying the constraint (\ref{CG=0}) such that 
\begin{eqnarray}
& & \hspace{-1cm} g_I^\varepsilon \to G_I, \quad \forall (x,c) \in \Omega \times {\mathbb R}^3
,
\label{IClim} 
\end{eqnarray}
as $\varepsilon \to 0$ and we prescribe the boundary condition to be
\begin{eqnarray}
& & \hspace{-1cm} G|_{t=0} =  G_I. 
,
\label{IClimit} 
\end{eqnarray}

\subsubsection{Comparison with the literature}
\label{subsub_compar}

Let us introduce the following classical drift velocities due to the variations
of the electromagnetic field
\cite[Chapter~4]{haz_mei_03}:
\begin{eqnarray*}
& & V_{cd} = \frac{c_{||}^2}{|B|} {\mathbf f}, \quad \text{the curvature drift due
to the magnetic curvature}\;(b \cdot \nabla_x) b,  \\
& & V_{gd} = \frac{\mu}{|B|} (b \times \nabla_x |B|), \quad \text{the gradient drift,}\\
& & V_{ed} = \frac{1}{|B|} (E \times b),\quad \text{the electric drift,} \\
& & V_{d||}= \mu \big( b \cdot (\nabla_x \times b) \big) b, \quad \text{a drift
parallel to the magnetic field}.  
\end{eqnarray*}
With these definitions, the operator ${\mathcal S}^\dagger$ takes the following form
\begin{eqnarray*}
& &  \hspace{-1cm} {\mathcal S}^\dagger   {\mathcal G} :=  \nabla_x \cdot
\left[\big( u_{||}b + V_{d||} + V_{ed} + V_{cd} + V_{gd} \big) {\mathcal G}\right] \\
& & \hspace{0cm} + \frac{\partial}{\partial c_\parallel} \left[ \left( -b \cdot \nabla_x
u_\parallel - \nabla_x \cdot(V_{gd} + V_{d||}) + (b\cdot\nabla_x)b \cdot V_{ed} +
\frac{\nabla_x|B|}{|B|} \cdot (V_{d||} - V_{gd})\right) \, c_\parallel \,
{\mathcal G} \right]  \\
& & \hspace{0cm} + \frac{\partial}{\partial \mu} \left[ \left( -
\frac{\partial_t|B|}{|B|} - (b\cdot\nabla_x)b \cdot V_{ed} - \nabla_x\cdot V_{ed}
\right. \right.  \\
& & \hspace{5.73cm} \left. \left. + \frac{c_\parallel^2}{\mu|B|} \nabla_x\cdot V_{d||} -
\frac{\nabla_x|B|}{|B|} \cdot (V_{ed} + V_{d||})  \right)\mu \, {\mathcal G} \right].
\end{eqnarray*}

We recover the classical expressions of the spatial drift which can be found in the literature \cite[Chapter~4]{haz_mei_03}. By contrast, the drifts in $c_\parallel$ or $\mu$ which appear here do not appear in general in the literature. This work shows that, at least in the regime described by the proposed scaling, such drifts must be included otherwise, the limit model is not consistant with the Vlasov equation which we used as a starting point.

%%%%%%%%%%%%%%%%%%%%%%%%%%%%%%%%%%%%%%%%%%%%%%%%%%%%%%%%%%%%%%%
%%%%%%%%%%%%%%%%%%%%%%%%%%%%%%%%%%%%%%%%%%%%%%%%%%%%%%%%%%%%%%%
%%%%%%%%%%%%%%%%%%%%%%%%%%%%%%%%%%%%%%%%%%%%%%%%%%%%%%%%%%%%%%%
%%%%%%%%%%%%%%%%%%%%%%%%%%%%%%%%%%%%%%%%%%%%%%%%%%%%%%%%%%%%%%%
%%%%%%%%%%%%%%%%%%%%%%%%%%%%%%%%%%%%%%%%%%%%%%%%%%%%%%%%%%%%%%%
\setcounter{equation}{0}
\section{The asymptotic limit $\varepsilon \to 0$: preliminaries}
\label{sec_drift_kinetic_limit_prelim}

%%%%%%%%%%%%%%%%%%%%%%%%%%%%%%%%%%%%%%%%%%%%%%%%%%%%%%%%%%%%%%%

\subsection{The Hilbert expansion}
\label{subsec_Hilbert}

Our goal is to find the asymptotic limit 
$\varepsilon \to 0$ of
\eqref{V_varepsilon_g}, \eqref{M_varepsilon_u}. We start by assuming that $g^\varepsilon$
and $u^\varepsilon$ admit Hilbert expansions: 
$$g^\varepsilon = g_0 + \varepsilon g_1 + \varepsilon^2 g_2 + \ldots \quad \text{and}\quad
u^\varepsilon = u_0 + \varepsilon u_1 + \varepsilon^2 u_2 + \ldots $$

Inserting these expansions in equation \eqref{V_varepsilon_g}, we find :
\begin{eqnarray}
\varepsilon^{-2} \text{ term:} & \quad  \quad 
- (c\times B)\cdot\nabla_cg_0 & = 0,\label{coef-2}\\
\varepsilon^{-1} \text{ term:} &  \quad  \quad  -(c\times B)\cdot
\nabla_cg_1 & = c\cdot\nabla_xg_0 + {\mathbb F}_0
\cdot \nabla_c g_0,
\label{coef-1}\\
\varepsilon^{0} \text{ term:} &  \quad  \quad -(c\times B)\cdot
\nabla_cg_2 & = \frac{\partial g_0}{\partial t} + u_0 \cdot \nabla_x g_0 -
c \cdot ( (\nabla_x u_0) \nabla_c g_0 )  + \nonumber\\
& & \hspace{1.5cm} + c \cdot \nabla_x g_1 + {\mathbb F}_0
\cdot \nabla_c g_1 + {\mathbb F}_1
\cdot \nabla_c g_0 .\label{coef0}
\end{eqnarray}
with 
$${\mathbb F}_0= \frac{\nabla_x \cdot {\mathbb P}_0}{n_0}, \quad {\mathbb F}_1 = \left( \frac{\nabla_x \cdot {\mathbb P}}{n} \right)_1. $$
Similarly, in \eqref{M_varepsilon_u}, we obtain: 
\begin{eqnarray*}
\varepsilon^{-2} \text{ term:} & \quad  \quad \displaystyle {\mathbb F}_0=  &  E + u_0 \times B,  \\
\varepsilon^{-1} \text{ term:} & \quad  \quad \displaystyle  {\mathbb F}_1 =  &  u_1 \times B, 
\end{eqnarray*}
We will see that the following orders in the expansion of eqs. \eqref{V_varepsilon_g} and (\ref{M_varepsilon_u}) are not required.  

Next, we introduce some formal notations. We denote by $L$, $T$, $A$, and $D$ the following operators:
\begin{eqnarray}
& & Lg = -(c \times B) \cdot \nabla_c g,\label{def_L}\\
& & Tg = c \cdot \nabla_x g + {\mathbb F}_0 \cdot \nabla_c g, \label{def_T}\\
& & Ag = \displaystyle{\frac{\partial g}{\partial t}} + u_0 \cdot \nabla_x g - c \cdot ((\nabla_x u_0) \nabla_c g),  \label{def_A} \\
& & D g = {\mathbb F}_1
\cdot \nabla_c g. \label{def_Phi}
\end{eqnarray}
Then, eqs. \eqref{coef-2}--\eqref{coef0} take the form 
\begin{align}
& L g_0 = 0, \label{new:coef-2}\\
& L g_1 = T g_0 , \label{new:coef-1}\\
& L g_2 = A g_0 +Tg_1 + D g_0 . \label{new:coef0}
\end{align}

Since the leading order term in the expansion involves the operator $L$, which describes the effect of a circular motion around the magnetic field
lines at infinite angular velocity, we now specifically examine the properties of this operator. 
%%%%%%%%%%%%%%%%%%%%%%%%%%%%%%%%%%%%%%%%%%%%%%%%%%%%%%%%%%%%%%%
\subsection{Study of $L$} 
\label{subsub_oper_L}

We investigate the solutions of $Lg = 0$ and more generally, in a second step, of those of $Lg = h$ for a given $h$. Since $L$ operates on $c$ only, the $x$ and $t$ variables are omitted. In what follows, we will not seek precise statements about the functional spaces. We assume that the functions are as regular as needed for the following statements to be correct. We denote by $b = B/|B|$ the director of the magnetic field. We assume that it is always defined in the region of interest i.e. that $B$ does not vanish, and that it is as smooth as needed. 

We first introduce some notations. 
Let ${\mathbf e}_1$, ${\mathbf e}_2$ be two vectors such that the set $\{{\mathbf e}_1,{\mathbf e}_2,b\}$ forms a direct orthonormal basis of $\RR^3$ (i.e. $b={\mathbf e}_1\times {\mathbf e}_2$). We denote by $(c_1,c_2,c_3)$ the coordinates of $c$ in this basis, with $c_3 = c_\parallel = c \cdot b$. We denote by $c_\bot = c - c_\parallel b$ the normal component of $c$ to $b$. It has coordinates $(c_1,c_2,0)$ in this basis and its norm is given by $|c_\bot| = (c_1^2 + c_2^2)^{1/2} = (2e - c_\parallel^2 )^{1/2}$ where $e= |c|^2/2$ is the energy. 

Now, we introduce a coordinate system for $c$ which is derived from the cylindrical coordinates. Any $c$ such that $c_\bot \not = 0$ is uniquely defined by the triple $(e,c_\parallel,\alpha)$ with $(e,c_\parallel) \in {\mathcal D}$ and $\alpha \in {\mathbb S}^1$ where 
\begin{equation}
{\mathcal D} = \{ (e,c_\parallel) \, | \, e \geq  0, \, \,  - \sqrt{2e} \leq c_\parallel \leq \sqrt{2e} \}, 
\label{S_defin}
\end{equation}
and ${\mathbb S}^1$ is the one-dimensional torus ${\mathbb R}/(2 \pi {\mathbb Z})$, such that
\begin{eqnarray}
& & e = \frac{1}{2} (c_1^2 + c_2^2 + c_3^2), \label{CV_e} \\ 
& & c_\parallel = c_3, \label{CV_cp} \\
& & \cos \alpha = c_1/(c_1^2 + c_2^2)^{1/2}, \quad \sin \alpha = c_2/(c_1^2 + c_2^2)^{1/2} , \label{CV_al}
\end{eqnarray}
or conversely
\begin{eqnarray}
& & c_1 = (2e - c_\parallel^2 )^{1/2} \cos \alpha, \label{CV_c1} \\
& & c_2 = (2e - c_\parallel^2 )^{1/2} \sin \alpha, \label{CV_c2} \\
& & c_3 = c_\parallel .  \label{CV_c3} 
\end{eqnarray}
The angle $\alpha$ is the gyrophase while $\co$ is proportional to the gyroradius of the particle in the magnetic field. 

Let $g(c) = \tilde g (e, c_\parallel, \alpha)$ be the expression of $g$ in this coordinate system where the function $\tilde g (e, c_\parallel, \alpha)$ is defined on ${\mathcal D} \times {\mathbb S^1}$. Saying that $\alpha \in {\mathbb S}^1$ means that $\tilde g$ is $2 \pi$-periodic with respect to $\alpha$. 

\begin{lemma}
The null space $\ker L$ of $L$ consists of functions which only depend on the parallel component $c_\parallel = c \cdot b$ and on the energy $e = |c|^2/2$, i.e.
\begin{equation}
Lg = 0 \, \Longleftrightarrow \, \exists G (e,c_\parallel) \, \mbox{ with } \, (e,c_\parallel) \in {\mathcal D} , \, \mbox{ such that } \, g(c) = G (|c|^2/2 \, , \, c \cdot b). 
\label{null_space_L}
\end{equation}
\label{lem_nullspace_L}
\end{lemma}
\noindent
{\bf Proof:} By elementary algebra, we have
$$\nabla_c g= \frac{\partial \tilde g}{\partial e} \, c 
 + \frac{\partial \tilde g}{\partial c_\parallel} \, b +
\frac{\partial \tilde g}{\partial \alpha} \, \frac{b \times c}{2e - c_\parallel^2},$$
and therefore, taking into account that $|b\times c|=\co$, we obtain
\begin{eqnarray}\label{L=deriv_alpha}
\widetilde{Lg}=|B|\, \frac{\partial \tilde g}{\partial \alpha}. 
\end{eqnarray}
Hence, $Lg = 0$ implies that $\partial \tilde g/\partial \alpha=0$, i.e.
$$\tilde g = \tilde g (e, c_\parallel),$$
which proves (\ref{null_space_L}). 
\endproof

In order to solve eq. $Lg=h$, we introduce the gyroaveraging operator
$\Pi$ defined for every function $g(c)$ by
\begin{equation}\label{def_gyromoy}
\Pi g\,(\underline e,\underline{c_\parallel})=\frac1{2\pi}\int_{{\mathbb R}^3}  g(c)
\,\delta\left(\frac{|c|^2}{2} - \underline e\right)\,\delta(c \cdot b-\underline{c_\parallel}) \,dc = \frac1{2\pi}\int_{{\mathbb S}^1} \tilde g(\underline e,\underline{c_\parallel},\alpha) \,d\alpha,
\end{equation}
for all $(\underline e,\underline{c_\parallel})\in{\mathcal D}$. $\Pi g$ is nothing but the
mean value of $\tilde g$ over the phase $\alpha$.
%We now define a projection operator onto phase-independent functions by 
%\begin{equation} P h (c)=\Pi h \, (\frac{|c|^2}{2}, c \cdot b). \label{def_P} \end{equation}

Since our asymptotic model will be obtained by gyroaveraging the system \eqref{new:coef-2}-\eqref{new:coef0}, the following properties of $\Pi$ will be useful. Their proof is easy and is left to the reader: 
\begin{lemma}
\label{lem_prop_Pi}
For any function $g(x,c,t)$, we have:
\begin{eqnarray}
& & \Pi \left( \frac{\partial g}{\partial x_i} \right) =  \frac{\partial}{\partial x_i} (\Pi g) + \frac{\partial b}{\partial x_i} \cdot \frac{\partial }{\partial c_\parallel} ( \Pi(cg) ) \, ,\label{dxPi} \\ 
& & \Pi \left( \frac{\partial g}{\partial t} \right) = \frac{\partial}{\partial t} (\Pi g) + \frac{\partial b}{\partial t} \cdot \frac{\partial }{\partial c_\parallel} ( \Pi(cg) ) \, ,\label{dtPi} \\ 
& & \Pi \left( \frac{\partial g}{\partial c_i} \right) =  \frac{\partial }{\partial e} ( \Pi (gc_i) ) +
 b_i \frac{\partial }{\partial c_\parallel} (\Pi g) \, , \label{pinablac}
\end{eqnarray}
\end{lemma}

We deduce the following properties, which will be useful in the sequel: 
\begin{lemma}
\label{lem_prop_Pi_2}
We have:
\begin{eqnarray}
& & \Pi  L=0 , \label{PL} \\
& & f \in \ker L \quad  \Longleftrightarrow \quad \tilde f = \Pi f, \label{kerL} 
\end{eqnarray}
and, for any function $g$, 
\begin{eqnarray}
& & \Pi T g = \left( \nabla_x + {\mathbb F}_0 \frac{\partial }{\partial e} \right) \cdot \Pi (cg) + (\nabla_x b) : \frac{\partial }{\partial c_\parallel} \Pi (c \otimes c \, g) 
+ ({\mathbb F}_0 \cdot b) \frac{\partial }{\partial c_\parallel} \Pi g,
\label{PiT} \\
& & \Pi A g = \left( \frac{\partial }{\partial t} + u_0 \cdot \nabla_x \right) \Pi g + \left( \left(  \frac{\partial }{\partial t} + u_0 \cdot \nabla_x \right) b \right) \cdot \frac{\partial }{\partial c_\parallel} \Pi (c \, g) \nonumber \\
& & \hspace{1cm} - \nabla_x u_0 : \frac{\partial }{\partial e} \Pi (c \otimes c \, g)  - ( (\nabla_x u_0) b ) \cdot \frac{\partial }{\partial c_\parallel} \Pi (c \, g) 
+ (\nabla_x \cdot u_0) \Pi g
, \label{PiA} \\
& & \Pi D g =  {\mathbb F}_1 \cdot \frac{\partial }{\partial e}  \Pi (c \, g)  +
( {\mathbb F}_1 \cdot b ) \frac{\partial }{\partial c_\parallel} \Pi g, \label{PiPhi}
\end{eqnarray}
where $:$ denotes the contracted product of two tensors.
\end{lemma}
\noindent
{\bf Proof:} (\ref{PL}) is immediately deduced from (\ref{pinablac}) and the fact that $(B \times c) \cdot c = (B \times c) \cdot b = 0$. (\ref{kerL}) is obvious from the definition (\ref{def_gyromoy}) and Lemma \ref{lem_nullspace_L}. The other formulas are simple applications of Lemma \ref{lem_prop_Pi}. \endproof

Concerning the moments of $g$, we have the 
\begin{lemma}
For any function $g(x,c,t)$, we have: 
\begin{eqnarray*}
& & n = \int_{{\mathbb R}^3} g \, dc = \int_{{\mathcal D}} \Pi g (e,c_\parallel) \, 2 \pi \, de \, dc_\parallel ,  \\
& & \int_{{\mathbb R}^3} g \, c_\parallel \, dc = \int_{{\mathcal D}} \Pi g (e,c_\parallel)\, c_\parallel \, 2 \pi \, de \, dc_\parallel . 
\end{eqnarray*}
Let $g$ be a function lying in $\ker L$ for all $(x,t)$, i.e. such that $g = G(x,\, |c|^2/2,\,c\cdot b,t)$. We have:
\begin{eqnarray}
& & n = \int_{{\mathcal D}} G (e,c_\parallel) \, 2 \pi \, de \, dc_\parallel , \label{densite_G} \\
& & {\mathbb P} = \int_{{\mathbb R}^3} g \, c \otimes c \, dc = p_\bot (\mbox{Id} - b \otimes b) + p_\parallel b \otimes b , \label{Pi_P} \\
& & p_\bot = \int_{{\mathcal D}}  G (e,c_\parallel) \, \left(e- \frac{1}{2} c_\parallel^2\right) \,  2 \pi \, de \, dc_\parallel, \quad  p_\parallel = \int_{{\mathcal D}}  G (e,c_\parallel) \,  c_\parallel^2 \,  2 \pi \, de \, dc_\parallel .\label{Pi_pressions} 
\end{eqnarray}
In particular, we have 
\begin{eqnarray}
& & \hspace{-1cm} \nabla_x \cdot {\mathbb P} = \nabla_x p_\bot + \Bigl[ b \cdot \nabla_x(p_\parallel - p_\bot) + (p_\parallel-p_\bot) (\nabla_x \cdot b)\Bigl] \, b + (p_\parallel-p_\bot) (b \cdot \nabla_x) b 
.\label{nabla_Pi_pressions} 
\end{eqnarray}
\label{lem_integ}
\end{lemma}
%%%%%%%%%%%%%
\begin{lemma}
Let $h$ be given. Then, equation
\begin{equation}
Lg = h,
\label{Lg=h}
\end{equation}
admits a solution if and only if $\tilde h$ has zero phase-average i.e. 
\begin{equation}
\Pi h = 0.
\label{solv}
\end{equation}
If the solvability condition (\ref{solv}) is satisfied, all solutions of (\ref{Lg=h}) are written
\begin{equation}
\tilde g (e, c_\parallel, \alpha) = \frac{1}{|B|} \,  \int_0^\alpha \tilde h (e, c_\parallel, \varphi) \, d \varphi + K(e,c_\parallel),
\label{sol_Lg=h}
\end{equation}
where $K$ is arbitrary. $K$ can be uniquely determined if we impose to $g$ to satisfy the cancellation condition 
\begin{equation}
\Pi g = 0.
\label{cancel}
\end{equation}
The so-defined unique solution of (\ref{Lg=h}) is denoted by $g = L^{-1} h$ and the operator $L^{-1}$ is called the pseudo-inverse of $L$. We have
\begin{equation}
\widetilde{L^{-1}h} (e, c_\parallel, \alpha) = \frac{1}{|B|} \,  \int_0^{2 \pi} \Gamma(\alpha, \varphi) \, \tilde h (e,c_\parallel, \varphi) \, d \varphi ,
\label{L-1h}
\end{equation}
where the Green kernel $\Gamma(\alpha, \varphi)$ is given by
\begin{equation}
\Gamma(\alpha, \varphi) = \left\{ \begin{array}{lll} 
\displaystyle \frac{\varphi}{2 \pi}, & \mbox{if} & 0 \leq \varphi < \alpha, \\
& & \\
\displaystyle \frac{\varphi}{2 \pi} - 1, & \mbox{if} & \alpha \leq \varphi \leq  2 \pi.
\end{array} \right.
\label{Gamma}
\end{equation}
$L^{-1} h $ does not depend on the particular choice of the basis vectors $({\mathbf e}_1,{\mathbf e}_2)$. 
If $B$ is a smooth function of $(x,t)$, $L^{-1}g$ is a smooth function of $(x,c,t)$ in the open set where $B$ does not vanish. \label{lem_Lg=h}
\end{lemma}
\noindent
{\bf Proof:} Acting $\Pi$ onto eq. (\ref{Lg=h}) and using (\ref{PL})
shows that (\ref{solv}) is a necessary condition. Conversely, using (\ref{L=deriv_alpha}), we see that any solution of (\ref{Lg=h}) is of the form (\ref{sol_Lg=h}). The only thing to prove is that  
this formula provides a periodic function of $\alpha$ (otherwise it is not possible to invert the change to cylindrical coordinates). This is true precisely if and only if $h$ satisfies the solvability condition (\ref{solv}). Applying the cancellation condition (\ref{cancel}) allows to uniquely determine $K$: 
$$ K(e,c_\parallel) =  - \frac{1}{2 \pi |B|} \,  \int_0^{2 \pi} \tilde h (e,c_\parallel, \varphi) \, (2 \pi - \varphi) \,  d \varphi \,.$$
Inserting this expression into (\ref{sol_Lg=h}), we find the expression of $L^{-1} h$ given by (\ref{L-1h}), 
(\ref{Gamma}).
We note that $\Gamma$ can be extended by periodicity into a function of $\varphi$ defined on ${\mathbb S}^1$. Now, if we add any constant to $\Gamma$, we do not change the result of (\ref{L-1h}) due to the solvability constraint (\ref{solv}). Precisely, it is an easy matter to see that, changing the origin ${\mathbf e}_1$ from which we measure the phase $\alpha$ amounts to add a fixed constant to $\alpha$, or to add a fixed constant to $\Gamma$. This shows that, despite the use of an origin for the phase $\alpha$ in (\ref{L-1h}),  $L^{-1}$ actually does not depend on this choice. As a result, $L^{-1} h$ only depends on $b$ and has the same regularity as $b$, i.e. it is smooth in the domain where $B$ is smooth and non-zero. 
\endproof

\begin{remark}
The choice of the vectors ${\mathbf e}_1,{\mathbf e}_2$ in the proofs above can be arbitrary. In particular, they are {\bf not} required to form a smooth vector field. We note that this choice is equivalent to choose an origin for the phase $\alpha$. By contrast, $c_\parallel$ and $e$ do not depend on ${\mathbf e}_1,{\mathbf e}_2$. We have shown that the pseudo-inverse $L^{-1}$, despite the fact that its analytical expression (\ref{L-1h}) looks dependent on the choice of a particular origin for the phase, actually does not depend on it. This remark is important because generating a smooth orthogonal basis ${\mathbf e}_1,{\mathbf e}_2$ in the plane normal to $b$ is always locally possible but may be globally difficult. Additionally, such a basis is not always associated to an admissible coordinate system. Indeed, for this to be possible, commutation relations are required. This obviously is not the case in general. 
\end{remark}
%%%%%%%%%%%%%%%%%%%%%%%%%%%%%%%%%%%%%%%%%%%%%%%%%%%%%%%%%%%%%%%
%%%%%%%%%%%%%%%%%%%%%%%%%%%%%%%%%%%%%%%%%%%%%%%%%%%%%%%%%%%%%%%
%%%%%%%%%%%%%%%%%%%%%%%%%%%%%%%%%%%%%%%%%%%%%%%%%%%%%%%%%%%%%%%
%%%%%%%%%%%%%%%%%%%%%%%%%%%%%%%%%%%%%%%%%%%%%%%%%%%%%%%%%%%%%%%
%%%%%%%%%%%%%%%%%%%%%%%%%%%%%%%%%%%%%%%%%%%%%%%%%%%%%%%%%%%%%%%
\setcounter{equation}{0}
\section{The asymptotic model: derivation}
\label{sec_gyro_model}
%%%%%%%%%%%%%%%%%%%%%%%%%%%%%%%%%%%%%%%%%%%%%%%%%%%%%%%%%%%%%%%
\subsection{The asymptotic model in abstract form}
\label{subsec_gyro_model_abstract}
In this section we derive an asymptotic model for the limit $g_0$ of $g^\varepsilon$ by formally passing to the limit
$\varepsilon\to0$ in \eqref{V_varepsilon_g}-\eqref{M_varepsilon_u}.  This model will be deduced by solving the sequence of equations appearing in the Hilbert expansion \eqref{new:coef-2}-\eqref{new:coef0}. First, by a simple application of Lemma \ref{lem_nullspace_L}, eq. (\ref{new:coef-2}) is easily solved by:

\begin{proposition}\label{propo:1}%[formal]
There exists a function $G(x,e, c_\parallel,t)$ such that
\begin{equation}
g_0(x,c,t)=G(x,|c|^2/2 \, , \, c \cdot b,t). \label{ordre_0}
\end{equation}
\end{proposition}

Now, the goal is to find the equation satisfied by $g_0$ or $G$. For this purpose, we turn 
to~(\ref{new:coef-1}). By a simple application of Lemma \ref{lem_Lg=h}, we have: 

\begin{proposition}
\label{propo:2}
Eq. \eqref{new:coef-1} admits a solution $g_1$ if and only if $g_0$ satisfies the solvability condition 
$$\Pi T g_0 = 0. $$
If this condition is satisfied, there exists a function $k \in \ker L$ (in other words, there exists a function $K(x,e, c_\parallel,t)$ and $k = K(x,|c|^2/2 \, , \, c \cdot b,t)$) such that
$$g_1=L^{-1} T g_0  + k.$$
\end{proposition}

Note that, following (\ref{contr}) which has been proven equivalent to (\ref{M_varepsilon_u}), we have:
\begin{eqnarray*}
\int g^1 \, c_\parallel \, dc = 0 .  
\end{eqnarray*}
Since, by construction, $\Pi L^{-1} = 0$, and $\int f \, dc = \int \Pi f \, dc$ for all functions $f$,  we can write:
\begin{eqnarray*}
\int L^{-1} T g^0 \, c_\parallel \, dc = \int \Pi L^{-1} T g^0 \, c_\parallel \, dc = 0 .  
\end{eqnarray*}
Therefore, we deduce that 
\begin{eqnarray*}
\int k \, c_\parallel \, dc = 0 .  
\end{eqnarray*}

Finally, again, by a simple application of Lemma \ref{lem_Lg=h}, the equation satisfied by $g_0$ appears as the solvability condition of (\ref{new:coef0}). For such a function, we compute 

\begin{proposition}
\label{propo:3}
Eq. \eqref{new:coef0} admits a solution $g_2$ if and only if $g_0$ satisfies the equation 
$$\Pi \big( A g_0 + T g_1 + D g_0 \big) =0.$$
\end{proposition}

We now collect the model in the following theorem and discuss its properties. 
\begin{theorem}
The formal limit $\varepsilon \to 0$ of problem (\ref{V_varepsilon_g}), (\ref{M_varepsilon_u}) leads to 
$g^{\eps}\rightarrow g_0$ and $u^{\eps}\rightarrow u_0$, where $g_0$ and $u_0$ are solutions to the following
abstract model: there exists a function $k$ such that 
\begin{eqnarray}
& & \tilde g_0 = \Pi g_0 , \quad \tilde k = \Pi k, \quad \mbox{(i.e. } \tilde g_0 \mbox{ and } \tilde k \mbox{ are independent of } \alpha \mbox{ )}, 
\label{eq:gyro_1} \\ 
& & \Pi\big(A g_0+TL^{-1}T g_0 + Tk +D g_0\big)=0,\label{eq:gyro_2}\\
& & \Pi(T g_0)=0\label{PiT_tgo=0},\\
& & {\mathbb F}_0 = \frac{\nabla_x \cdot {\mathbb P}_0}{n_0} = E + u_0 \times B, \quad {\mathbb F}_1 = \left( \frac{\nabla_x \cdot {\mathbb P}}{n} \right)_1 =   u_1 \times B, \label{eq:gyro_3} \\
& & \int k \, c_\parallel \, dc = 0 ,  
\label{intkcpar=0}
\end{eqnarray}
where we recall that the tilde means that function is expressed in the coordinate system (\ref{CV_e})-(\ref{CV_al}) and where the operators $\Pi$, $A$, $T$, $L^{-1}$ and $D$ are respectively defined 
by \eqref{def_gyromoy}, \eqref{def_A}, \eqref{def_T}, \eqref{L-1h} and \eqref{def_Phi}.
\label{thm_abstract_gyro}
\end{theorem}

Let us first note that eqs. \eqref{eq:gyro_1} means that $\tilde g_0$ and $\tilde k$ do not depend 
of the gyrophase $\alpha$.
In this model, $g_0$ is determined by eq. \eqref{eq:gyro_2} while the unknown function $k$ plays the role 
of the Lagrange multiplier associated to the constraint \eqref{PiT_tgo=0}. 
We will see in section~\ref{explicit_eq_for_u} that $u_0$ is fully determined by the first equation of \eqref{eq:gyro_3}. Finally, it is not necessary to determine $g_1$ and $u_1$. Indeed, $g_1$ appears
in the definition of $D$ but in the next section, using~\eqref{eq:gyro_3}, we prove that $\Pi D g_0=0$.

In the next sections, we make this model explicit in terms of a partial differential system.

%%%%%%%%%%%%%%%%%%%%%%%%%%%%%%%%%%%%%%%%%%%%%%%%%%%%%%%%%%%%%%%
%%%%%%%%%%%%%%%%%%%%%%%%%%%%%%%%%%%%%%%%%%%%%%%%%%%%%%%%%%%%%%%
\subsection{The asymptotic model: explicit form}
\label{subsec_gyro_model_explicit}

%%%%%%%%%%%%%%%%%%%%%%%%%%%%%%%%%%%%%%%%%%%%%%%%%%%
\subsubsection{The constraint $\Pi T g_0 = 0$}
\label{subsub_PiTg0=0}

In this section, we consider the constraint (\ref{PiT_tgo=0}): 

\begin{proposition}
\label{propo:2_suite}
Condition (\ref{PiT_tgo=0}) is equivalent to the following equation for $G$: 
$$\left( \nabla_x + {\mathbb F}_0 \frac{\partial}{\partial e} \right) \cdot ( c_\parallel G b) + (\nabla_x \cdot b) \frac{\partial}{\partial c_\parallel} \left( ( e - \frac{1}{2}c_\parallel^2 ) \, G \right)  + ( {\mathbb F}_0 \cdot b ) \frac{\partial G}{\partial c_\parallel} = 0. $$
\end{proposition}
\noindent
{\bf Proof:} We apply (\ref{PiT}) with $g = g_0$ of the form (\ref{ordre_0}). For such a function, an easy computation shows that 
\begin{eqnarray}
& & \Pi (c\, g) = c_\parallel G b, \label{Picg}\\
& & \Pi (c \otimes c \, g) = G \, \left[ ( e - \frac{1}{2}c_\parallel^2 ) (\mbox{Id} - b \otimes b ) + c_\parallel^2  b \otimes b \right].
\label{Picocg}
\end{eqnarray}
Then, we insert (\ref{Picg}), (\ref{Picocg}) into (\ref{PiT}) by noticing that $(\nabla_x b) : (b \otimes b) = ((b \cdot \nabla_x) b) \cdot b = 0$, since $|b|=1$, and that $(\nabla_x b) : \mbox{Id} = \nabla_x \cdot b$. \endproof
%%%%%%%%%%%%%%%%%%%%%%%%%%%%%%%%%%%%%%%%%%%%%%%%%%%
\subsubsection{The main equation (\ref{eq:gyro_2})}
\label{subsub_gyro2}

We start with $\Pi A g_0$. 

\begin{lemma}
We have: 
\begin{eqnarray}
& & \Pi A g_0 = \frac{\partial G}{\partial t} 
+ \nabla_x \cdot (u_0 G) - (\nabla_x \cdot u_0) \, \frac{\partial }{\partial e} \left[ (e - \frac{1}{2} c_\parallel^2) G \right] \nonumber \\
& & \hspace{3cm}  - b \cdot ((\nabla_x u_0) b) \, 
\left\{ 
\frac{\partial }{\partial e} \left[ (-e + \frac{3}{2} c_\parallel^2) G \right] 
+ \frac{\partial }{\partial c_\parallel} (c_\parallel G) 
\right\} .
\label{PiAg_0}
\end{eqnarray}
\label{lem_PiAg_0}
\end{lemma}
\noindent
{\bf Proof:} 
We insert (\ref{Picg}), (\ref{Picocg}) into (\ref{PiA}) and get
\begin{eqnarray}
& & \Pi A g_0 = \left( \frac{\partial }{\partial t} + u_0 \cdot \nabla_x \right) G + \left( \left(  \frac{\partial }{\partial t} + u_0 \cdot \nabla_x \right) b \right) \cdot \frac{\partial }{\partial c_\parallel} (c_\parallel G b)  \nonumber \\
& & \hspace{3cm} - \nabla_x u_0 : \frac{\partial }{\partial e} \left\{ G \, \left[ ( e - \frac{1}{2}c_\parallel^2 ) (\mbox{Id} - b \otimes b ) + c_\parallel^2  b \otimes b \right] \right\} \nonumber \\
& & \hspace{3cm}  - ( (\nabla_x u_0) b ) \cdot \frac{\partial }{\partial c_\parallel} (c_\parallel G b) 
+ (\nabla_x \cdot u_0) G. 
\label{PiA_g0} 
\end{eqnarray}
We first note that 
$$ \left( \left(  \frac{\partial }{\partial t} + u_0 \cdot \nabla_x \right) b \right) \cdot b = 0 , $$
since $|b|=1$, which implies that the second term of (\ref{PiA_g0}) vanishes. Then, noting that $(\nabla_x u_0):\mbox{Id} = \nabla_x \cdot u_0$ and $(\nabla_x u_0):(b \otimes b) = ((\nabla_x u_0) b) \cdot b$, the other terms combine into~(\ref{PiAg_0}). \endproof

We now denote $\gamma_1 = L^{-1} T g_0$. With this definition, we have $g_1 = \gamma_1+k$. We now compute $\Pi T \gamma_1$. By inspection of (\ref{PiT}), we realize that we need to compute $\Pi (c \, \gamma_1)$ and $\Pi ((c \otimes c) \, \gamma_1)$ (we note that because of (\ref{cancel}), $\Pi \gamma_1 = 0$). We first compute $\Pi (c \, \gamma_1)$: 

\begin{lemma}
We have:
\begin{eqnarray}
& & \Pi (c \, \gamma_1) = \frac{1}{|B|} (e - \frac{1}{2} c_\parallel^2) \, \, \,   b \times \left[ \left( \nabla_x  + {\mathbb F}_0 \frac{\partial }{\partial e}\right) G + c_\parallel \frac{\partial G}{\partial c_\parallel} (b \cdot \nabla_x) b \right] 
. 
\label{Picgamma1} 
\end{eqnarray}
\label{lem_Picgamma1}
\end{lemma}
\noindent
{\bf Proof:} 
postponed in the Appendix. \endproof

\begin{lemma}
We have:
\begin{eqnarray}
& & (\nabla_x b) : \frac{\partial}{\partial c_\parallel} \Pi ((c \otimes c) \, \gamma_1) = \nonumber \\
& & \hspace{2cm} = - \frac{1}{|B|} \frac{\partial}{\partial c_\parallel} \left[ (e - \frac{1}{2} c_\parallel^2) c_\parallel \, \, \,   \Big(b \times ((b \cdot \nabla_x) b) \Big) \cdot  \left( \nabla_x  + {\mathbb F}_0 \frac{\partial }{\partial e}\right) G  \right]
. 
\label{Picocgamma1} 
\end{eqnarray}
\label{lem_Picocgamma1}
\end{lemma}
\noindent
{\bf Proof:}  postponed in the Appendix.\endproof

By collecting formulas (\ref{Picgamma1}) and (\ref{Picocgamma1}) and inserting them into (\ref{PiT}), we obtain the 

\begin{lemma}
We have: 
\begin{eqnarray*}
& &  \hspace{-1cm}\Pi T \gamma_1 = \left( \nabla_x + {\mathbb F}_0 \frac{\partial}{\partial e} \right) \cdot \left\{
\frac{1}{|B|} (e - \frac{1}{2} c_\parallel^2) \, \, \,   b \times \left[ \left( \nabla_x  + {\mathbb F}_0 \frac{\partial }{\partial e}\right) G 
 + c_\parallel \frac{\partial G}{\partial c_\parallel} (b \cdot \nabla_x) b \right] \right\} \\
[8pt]& & \hspace{2.cm} - \frac{1}{|B|} \frac{\partial}{\partial c_\parallel} \left[ (e - \frac{1}{2} c_\parallel^2) c_\parallel \, \, \,   \Big(b \times ((b \cdot \nabla_x) b) \Big) \cdot  \left( \nabla_x  + {\mathbb F}_0 \frac{\partial }{\partial e}\right) G  \right].
\end{eqnarray*}
\label{lem_PiTgamma1}
\end{lemma}

Since $\partial\tilde k/\partial\alpha=0$, the computation of $\Pi T k$ is the same as for $\Pi T g_0$:

\begin{lemma}
We have:
$$\Pi T k = \left( \nabla_x + {\mathbb F}_0 \frac{\partial}{\partial e} \right) \cdot ( c_\parallel K b) + (\nabla_x \cdot b) \frac{\partial}{\partial c_\parallel} \left( ( e - \frac{1}{2}c_\parallel^2 ) \, K \right)  + ( {\mathbb F}_0 \cdot b ) \frac{\partial K}{\partial c_\parallel}. $$
\label{lem_PiTk}
\end{lemma}

Finally we calculate $\Pi D g_0$.

\begin{lemma}
We have:
$$\Pi D g_0  = 0. $$
\label{lem_PiPhig0}
\end{lemma}
\noindent
{\bf Proof:} Thanks to (\ref{def_Phi}), (\ref{pinablac}) and (\ref{Picg}), we get:
\begin{eqnarray*}
\Pi D g_0 & = & {\mathbb F}_1 \cdot \frac{\partial }{\partial e} ( \Pi (cg_0) ) +
 ({\mathbb F}_1 \cdot b) \frac{\partial }{\partial c_\parallel} (\Pi g_0), \\
 &=& ({\mathbb F}_1 \cdot b) \left( \frac{\partial }{\partial e} ( c_\parallel G ) +
\frac{\partial }{\partial c_\parallel} G \right).
\end{eqnarray*}
But, with (\ref{eq:gyro_3}), \, ${\mathbb F}_1 \cdot b = 0$. This concludes the proof. 
\endproof

\subsubsection{The explicit form of the asymptotic model}
From now on, for simplicity, we write $u$ and ${\mathbb F}$ instead of
$u_0$ and ${\mathbb F}_0$, respectively.
We collect all the previous lemmas in the following theorem, which gives the explicit form of the drift kinetic model: 

\begin{theorem}
The formal limit $\varepsilon \to 0$ of problem (\ref{V_varepsilon_g})-(\ref{M_varepsilon_u}) leads to 
$g^{\eps}\rightarrow g_0$ and $u^{\eps}\rightarrow u$. There exists $G$ such that
$$g_0(x,c,t)=\tilde g_0(x,e,c_{\parallel},\alpha,t)=G(x,e,c_{\parallel},t),$$
and $G$ and $u$ are solutions to 
the following partial differential system: 
\begin{eqnarray}
& & \hspace{-1cm}\frac{\partial G}{\partial t} 
+ \nabla_x \cdot (u G) - (\nabla_x \cdot u) \, \frac{\partial }{\partial e} \left[ (e - \frac{1}{2} c_\parallel^2) G \right] \nonumber \\
& & \hspace{3cm}  - b \cdot ((\nabla_x u) b) \, 
\left\{ 
\frac{\partial }{\partial e} \left[ (-e + \frac{3}{2} c_\parallel^2) G \right] 
+ \frac{\partial }{\partial c_\parallel} (c_\parallel G) 
\right\} \nonumber \\
& & \hspace{0.5cm}+ \left( \nabla_x + {\mathbb F} \frac{\partial}{\partial e} \right) \cdot \left\{
\frac{1}{|B|} (e - \frac{1}{2} c_\parallel^2) \, \, \,   b \times \left[ \left( \nabla_x  + {\mathbb F} \frac{\partial }{\partial e}\right) G 
+ c_\parallel \frac{\partial G}{\partial c_\parallel} (b \cdot \nabla_x) b \right] \right\} \nonumber \\
[5pt]& & \hspace{0.5cm} - \frac{1}{|B|} \frac{\partial}{\partial c_\parallel} \left[ (e - \frac{1}{2} c_\parallel^2) c_\parallel \, \, \,   \Big(b \times ((b \cdot \nabla_x) b) \Big) \cdot  \left( \nabla_x  + {\mathbb F} \frac{\partial }{\partial e}\right) G  \right] \nonumber \\
[5pt]& & \hspace{0.5cm}+  \left( \nabla_x + {\mathbb F} \frac{\partial}{\partial e} \right) \cdot ( c_\parallel K b) + (\nabla_x \cdot b) \frac{\partial}{\partial c_\parallel} \left( ( e - \frac{1}{2}c_\parallel^2 ) \, K \right)  + ( {\mathbb F} \cdot b ) \frac{\partial K}{\partial c_\parallel} =0 , \label{gyro_expl_1}\\
& & \nonumber \\
& & \hspace{-1cm}\left( \nabla_x + {\mathbb F} \frac{\partial}{\partial e} \right) \cdot ( c_\parallel G b) + (\nabla_x \cdot b) \frac{\partial}{\partial c_\parallel} \left( ( e - \frac{1}{2}c_\parallel^2 ) \, G \right)  + ( {\mathbb F} \cdot b ) \frac{\partial G}{\partial c_\parallel} = 0 , 
\label{gyro_expl_2} \\
& & \nonumber \\
& & \hspace{-1cm}{\mathbb F} = \frac{\nabla_x \cdot {\mathbb P}}{n} = E + u \times B,  \label{gyro_expl_3} \\
& & \int K \, c_\parallel \, dc = 0 ,  
\label{intKcpar=0}
\end{eqnarray}
where $K=K(x,e,c_\parallel,t)$ is completely determined by the constraint
\eqref{gyro_expl_2} under some suitable boundary conditions, and where we recall
the expressions of $n$ and $\nabla_x \cdot {\mathbb P}$ are given in
\eqref{densite_G} and \eqref{nabla_Pi_pressions} respectively.
\label{th_gyro_explicit}
\end{theorem}
In the following section, we write the model in terms of more relevant variables.

%%%%%%%%%%%%%%%%%%%%%%%%%%%%%%%%%%%%%%%%%%%%%%%%%%%%%%%%%%%%%%%
%%%%%%%%%%%%%%%%%%%%%%%%%%%%%%%%%%%%%%%%%%%%%%%%%%%%%%%%%%%%%%%
%%%%%%%%%%%%%%%%%%%%%%%%%%%%%%%%%%%%%%%%%%%%%%%%%%%%%%%%%%%%%%%
%%%%%%%%%%%%%%%%%%%%%%%%%%%%%%%%%%%%%%%%%%%%%%%%%%%%%%%%%%%%%%%
%%%%%%%%%%%%%%%%%%%%%%%%%%%%%%%%%%%%%%%%%%%%%%%%%%%%%%%%%%%%%%%
\subsection{The asymptotic model in the magnetic moment variable}
\label{sec_gyro_interpretation}

In order to highlight the physical relevance of the model, it is useful to introduce 
the new variable (see \cite{haz_ware_78})
$$\mu = \frac{1}{|B|} \Big(e - \frac{c_\parallel^2}{2}\Big).$$
The quantity $\mu$ is the magnetic moment of the particle in its rotation motion
about the magnetic field, see section~\ref{sec_main}. 

We introduce the change of variables 
$$ G(x,e,c_\parallel,t) = \overline G (x,\mu, c_\parallel, t) , \quad 
K(x,e,c_\parallel,t) = \overline K (x,\mu, c_\parallel, t)\,  . $$
We note that 
\begin{eqnarray}
(e,c_\parallel) \in {\mathcal D} \Longleftrightarrow (\mu,c_\parallel) \in \overline {\mathcal D} = {\mathbb R}_+ \times {\mathbb R}   , 
\label{chg_var_dom} 
\end{eqnarray}
where ${\mathcal D}$ is defined by (\ref{S_defin}). 

Since $B$ is divergence free, we have:
$$ \nabla_x \cdot b = B \cdot \nabla_x \left(\frac{1}{|B|}\right)  = - b \cdot \nabla_x (\ln |B| ).  $$ 
Furthermore, the following formulas hold true:
$$\begin{array}{ll}
\displaystyle\nabla_x G = \nabla_x \overline G - \mu \frac{\partial \overline G}{\partial \mu}\nabla_x(\ln |B|),
& \hspace{1cm}\displaystyle\frac{\partial G}{\partial e} = \frac{1}{|B|} \frac{\partial \overline G}{\partial \mu}  ,  \\
[10pt]\displaystyle\frac{\partial G}{\partial c_\parallel} = \frac{\partial \overline G}{\partial c_\parallel} - \frac{c_\parallel}{|B|} \frac{\partial \overline G}{\partial \mu}  , 
&\hspace{1cm}\displaystyle\frac{\partial G}{\partial t} = \frac{\partial \overline G}{\partial t} 
- \mu \frac{\partial \overline G}{\partial \mu} \, \frac{\partial}{\partial t}(\ln |B|).  
\end{array}$$
From now on, we work in the variables $(\mu, c_\parallel)$ and we drop the
overbars for clarity. From these formulas, we find

\begin{lemma}
In the new variables, the constraint (\ref{gyro_expl_2}) is written 
\begin{eqnarray}
& & {{\mathcal C}}  G :=  c_\parallel \, b \cdot \nabla_x  G + 
b \cdot \Big({\mathbb F} - \mu \nabla_x |B|\Big) \frac{\partial  G}{\partial c_\parallel} = 0 .
\label{tilde_C} 
\end{eqnarray}
A function $ G(x,\mu,c_\parallel,t)$ which satisfies the constraint (\ref{tilde_C}) is constant along the curves $(x=X(\tau),\mu={\mathcal M}(\tau),c_\parallel= C_\parallel(\tau))$ satisfying the  ODE system (\ref{dXdtau})-(\ref{dMdtau}). 
\label{lem_tilde_C}
\end{lemma}

\begin{lemma}
In the new variables, the asymptotic model (\ref{gyro_expl_1}), (\ref{gyro_expl_2}) is written 
\begin{eqnarray}
& & \hspace{-1cm} \frac{\partial  G}{\partial t} + {{\mathcal S}}  G + {{\mathcal C}}  K = 0 , \label{gyro_new_1} \\
& & {{\mathcal C}}  G = 0 . \label{gyro_new_2} 
\end{eqnarray}
where ${{\mathcal C}}$ is given by (\ref{tilde_C}) and ${{\mathcal S}}$ by
\begin{eqnarray}
& &  \hspace{-1cm} {{\mathcal S}}  G :=  \left(u + \mu \, \nabla_x \times b - b \times \Phi + (\frac{c_\parallel^2}{|B|} - \mu) \, {\mathbf f} \right) \cdot \nabla_x  G \nonumber \\
& & \hspace{0cm} \left(- (\nabla_x u) : (b \otimes b) + \mu \nabla_x \cdot {\mathbf f} + \Phi \cdot {\mathbf f} \phantom{\frac{1}{|B|}} \hspace{-0.6cm} \right) \, c_\parallel \frac{\partial  G}{\partial c_\parallel} \nonumber \\
& & \hspace{0cm} + \left[ - \left(\frac{\partial}{\partial t} + u \cdot \nabla_x \right)\ln |B| - \nabla_x \cdot u  + \nabla_x u : (b \otimes b) + \frac{1}{|B|} \nabla_x \cdot (B \times \Phi) \right. \nonumber \\
& & \hspace{8cm} \left. - \left(  \frac{c_\parallel^2}{|B|} \nabla_x \cdot {\mathbf f} + \Phi \cdot {\mathbf f} \right)
\right] \mu \frac{\partial  G}{\partial \mu} 
,
\label{tilde_S_2} 
\end{eqnarray}
with 
\begin{eqnarray*}
& & \Phi = \frac{1}{|B|} ({\mathbb F} - \mu \nabla_x |B|), \quad 
{\mathbf f} = b  \times (b \cdot \nabla_x) b,  \quad 
{\mathbb F} = \frac{\nabla_x \cdot {\mathbb P}}{n}, 
\end{eqnarray*}
and $n$ and ${\mathbb P}$ being given by  
\begin{eqnarray*}
& & n = \int_{\overline{\mathcal D}} G (\mu ,c_\parallel) \, 2 \pi |B| \, d \mu \, dc_\parallel, \quad {\mathbb P} = p_\bot (\mbox{Id} - b \otimes b) + p_\parallel b \otimes b ,  \\
& & p_\bot = \int_{\overline{\mathcal D}}  G (\mu ,c_\parallel) \, \mu |B| \,   2 \pi |B| \, d \mu \, dc_\parallel , \quad  p_\parallel = \int_{\overline{\mathcal D}} G (\mu, c_\parallel) \,  c_\parallel^2 \,   2 \pi |B| \, d \mu \, dc_\parallel  .
\end{eqnarray*}
Additionally, the velocity $u$ satisfies the constraint:
\begin{eqnarray}
& &  {\mathbb F} = E + u \times B. \label{vel_const}
\end{eqnarray}
and $K$ satisfies the constraint
\begin{eqnarray}
& & \int K \, c_\parallel \, d \mu \, dc_\parallel = 0 ,  
\label{intKcpar=0mu}
\end{eqnarray}
\label{lem_tilde_S}
\end{lemma}

\medskip
\noindent
{\bf Proof:} By performing the change of variables (\ref{chg_var_dom}) into (\ref{gyro_expl_1}), we readily find that 
\begin{eqnarray*}
& &  \hspace{0cm} {{\mathcal S}}  G :=  u \cdot \nabla_x  G - \nabla_x u : (b \otimes b) \, c_\parallel  \frac{\partial  G}{\partial c_\parallel} \nonumber \\
&  & \hspace{2.7cm} - \, \left( \left(\frac{\partial}{\partial t} + u \cdot \nabla_x \right)\ln |B| + \nabla_x \cdot u  - \nabla_x u : (b \otimes b) \right) \mu \frac{\partial  G}{\partial \mu} 
\nonumber \\
&  & \hspace{2.7cm} + \, \left(\nabla_x + \Phi \frac{\partial}{\partial \mu}\right) \cdot \left\{ \mu b \times  \left(\nabla_x + \Phi \frac{\partial}{\partial \mu}\right)  G \right\}  \\
& & \hspace{2.7cm} + \, \left(\nabla_x + \Phi \frac{\partial}{\partial \mu}\right) \cdot \left\{ \mu c_\parallel {\mathbf f} \, \left(\frac{\partial}{\partial c_\parallel} - \frac{c_\parallel}{|B|} \frac{\partial}{\partial \mu}\right)  G \right\}\\
& &  \hspace{5cm} 
-  \, \left(\frac{\partial}{\partial c_\parallel} - \frac{c_\parallel}{|B|} \frac{\partial}{\partial \mu}\right) \left\{ \mu c_\parallel {\mathbf f} \cdot \left(\nabla_x + \Phi \frac{\partial}{\partial \mu}\right)  G \right\}
.
\end{eqnarray*}
Now, we expand the second order operators and find: 
\begin{eqnarray*}
& &  \hspace{-1cm} 
 \left(\nabla_x + \Phi \frac{\partial}{\partial \mu}\right) \cdot \left\{ \mu b \times  \left(\nabla_x + \Phi \frac{\partial}{\partial \mu}\right)  G \right\} = 
 \left( \mu \, \nabla_x \times b - b \times \Phi  \phantom{\frac{1}{|B|}} \hspace{-0.6cm} \right) \cdot \nabla_x  G  \\
& & \hspace{8cm} + \left(  \frac{1}{|B|} \nabla_x \cdot (B \times \Phi) 
\right) \mu \frac{\partial  G}{\partial \mu} 
,
\end{eqnarray*}
and 
\begin{eqnarray*}
& &  \hspace{-1cm}  \left(\nabla_x + \Phi \frac{\partial}{\partial \mu}\right) \cdot \left\{ \mu c_\parallel {\mathbf f} \, \left(\frac{\partial}{\partial c_\parallel} - \frac{c_\parallel}{|B|} \frac{\partial}{\partial \mu}\right)  G \right\}
-  \, \left(\frac{\partial}{\partial c_\parallel} - \frac{c_\parallel}{|B|} \frac{\partial}{\partial \mu}\right) \left\{ \mu c_\parallel {\mathbf f} \cdot \left(\nabla_x + \Phi \frac{\partial}{\partial \mu}\right)  G \right\}  \\
& & \hspace{-0.5cm} = 
\left( \frac{c_\parallel^2}{|B|} - \mu  \right) \, {\mathbf f}  \cdot \nabla_x  G  + \left( \mu \nabla_x \cdot {\mathbf f} + \Phi \cdot {\mathbf f}  \phantom{\frac{1}{|B|}} \hspace{-0.6cm} \right) \, c_\parallel \frac{\partial  G}{\partial c_\parallel} - \left(  \frac{c_\parallel^2}{|B|} \nabla_x \cdot {\mathbf f} + \Phi \cdot {\mathbf f} 
\right) \mu \frac{\partial  G}{\partial \mu} , 
\end{eqnarray*}
where we have used that  
$$ \frac{\partial \Phi}{\partial \mu} = -\frac{\nabla_x |B|}{|B|}, $$
and that 
$$\nabla_x \cdot (b \times \Phi) - \Phi \cdot \left( b \times \frac{\nabla_x |B|}{|B|} \right) = \frac{1}{|B|} \nabla_x \cdot (B \times \Phi) .$$
Collecting these various formulas together leads to (\ref{tilde_S_2}).  \endproof

We obtain the conservative form of the model given in Theorem~\ref{main_result} defining
$${\mathcal G} = 2 \pi \, |B| \,  G, \quad {\mathcal K} = 2 \pi \, |B| \,  K.$$
Then, we have the

\begin{lemma}
The unknowns ${\mathcal G}$ and ${\mathcal K}$ satisfy system~(\ref{gyro_cons_1}),~(\ref{gyro_cons_2})
where ${\mathcal S}^\dagger$ is given by~(\ref{S_dagger}), 
 $\Phi$, ${\mathbf f}$ and ${\mathbb F}$ are defined by~(\ref{F_agag}), $n$ and ${\mathbb P}$ are given 
by~(\ref{Pi_P_again}),~(\ref{Pi_pressions_again}).
Finally the velocity satisfies the constraint  
\begin{eqnarray}
& &  {\mathbb F} = E + u \times B. \label{vel_const_again}
\end{eqnarray}
Additionally, ${\mathcal K}$ satisfies (\ref{Kflux=0}). 
\label{lem_var_cons}
\end{lemma}

The relations between ${\mathcal C}^\dagger$ and ${\mathcal S}^\dagger$ on the hand and ${\mathcal C}$ and ${\mathcal S}$ on the other hand, are clarified in section \ref{subsub_fast}.

%%%%%%%%%%%%%%%%%%%%%%%%%%%%%%%%%%%%%%%%%%%%%%%%%%%%%%%%%%%%%%%%%%
\subsection{Explicit equations for the limit fluid velocity}\label{explicit_eq_for_u}

To find explicit equations for the velocity, we will need the moments of ${\mathcal G}$. We introduce the general moments $M_{m,q}$
and $K_{m,q}$ of ${\mathcal G}$ and ${\mathcal K}$, which are defined by
(see \cite[Chapter~6]{haz_mei_03})
$$M_{m,q} = \int {\mathcal G} \,  c_\parallel^m \, \mu^q \,   d \mu \, dc_\parallel  , \quad   K_{m,q} = \int {\mathcal K} \,  c_\parallel^m \, \mu^q \,   d \mu \, dc_\parallel . $$
We first note that 
\begin{eqnarray}
& & n = M_{0,0} \, , \quad p_\parallel = M_{2,0} \, , \quad p_\bot = |B| M_{0,1}  . \label{nMKmq} 
\end{eqnarray}
Then, we obtain
\begin{lemma}
The moment system satisfied by $M_{m,q}$ and $K_{m,q}$ is as follows: 
\begin{eqnarray}
& &  \hspace{-1cm} \frac{\partial}{\partial t} M_{m,q} + 
\nabla_x \cdot \left[ \left(u - \frac{b \times {\mathbb F}}{|B|} \right) \, M_{m,q} + \left( \, |B| \, (\nabla_x \times  \frac{b}{|B|}) - {\mathbf f} \, \right) \, M_{m,q+1} + \frac{{\mathbf f}}{|B|} \, M_{m+2,q} \right] \nonumber \\
& & \hspace{0cm} + \left[ (m-q) \, (\nabla_x u) : (b \otimes b) - (m-q) \, \frac{{\mathbb F}}{|B|} \cdot {\mathbf f} + q \left(\frac{\partial}{\partial t} + u \cdot \nabla_x\right) \ln |B| \, +\right. \nonumber \\
& & \hspace{7cm} \left. + \,  q \, \nabla_x \cdot u - \frac{q}{|B|} \nabla_x \cdot (b \times {\mathbb F}) \right]  M_{m,q} \nonumber \\
& & \hspace{0cm} + \left[ - m \nabla_x \cdot {\mathbf f} + (m-q)\, \frac{\nabla_x |B|}{|B|} \cdot {\mathbf f} + \frac{q}{|B|} \nabla_x \cdot (b \times \nabla_x |B|) \right] M_{m,q+1} \nonumber \\
& & \hspace{0cm} + \left[ \frac{q}{|B|} \nabla_x \cdot {\mathbf f} \right]  M_{m+2,q} \nonumber \\
& & \hspace{0cm} + \nabla_x \cdot (K_{m+1,q} \, b) - m (b \cdot {\mathbb F}) K_{m-1,q} + m (b \cdot \nabla_x |B|) K_{m-1,q+1} = 0, \label{eq_Mmq} 
\end{eqnarray}
with the constraint
\begin{eqnarray}
& &  \hspace{-1cm} \nabla_x \cdot (M_{m+1,q} \, b) - m (b \cdot {\mathbb F}) M_{m-1,q} + m (b \cdot \nabla_x |B|) M_{m-1,q+1} = 0, \label{cont_Mmq}
\end{eqnarray}
and with the convention that any moment with negative indexes is identically zero. 
\label{lem_moments}
\end{lemma}

The infinite set of equations \eqref{eq_Mmq}-\eqref{cont_Mmq} for the positive integers
$m$ and $q$, contains the same information as the original model 
\eqref{gyro_cons_1}-\eqref{gyro_cons_2}.

In particular, the first moments lead to

\begin{lemma}
(i) The mass conservation equation holds: 
\begin{eqnarray}
&  & \hspace{-1cm} 
\frac{\partial n}{\partial t} + \nabla_x \cdot (nu) = 0
 \, . 
\label{Mass_cons} 
\end{eqnarray}

\noindent
(ii) We suppose that $M_{1,0} = 0$ at the boundary of the domain. Then,
\begin{eqnarray}
&  & \hspace{-1cm} 
M_{1,0} = \int  {\mathcal G} \, c_\parallel \, d\mu \, dc_\parallel = 0
 \, , 
\label{Int_G_cpar_0} 
\end{eqnarray}
everywhere.

\noindent
(iii) The constraint (\ref{cont_Mmq}) for $m=1$ and $q=0$ carries no information: it is redundant with the third equation in (\ref{F_agag}).

\noindent
(iv) The pressures satisfy the following equations: 
\begin{eqnarray}
& &  \hspace{-1cm} \frac{\partial p_\parallel}{\partial t}  + 
\nabla_x \cdot \left[ \left(u - \frac{b \times {\mathbb F}}{|B|} \right) \, p_\parallel + \left( \, |B| \, (\nabla_x \times  \frac{b}{|B|}) - {\mathbf f} \, \right) \, M_{2,1} + \frac{{\mathbf f}}{|B|} \, M_{4,0} \right] \nonumber \\
& & \hspace{0cm} + 2 \left[ \, (\nabla_x u) : (b \otimes b) -  \, \frac{{\mathbb F}}{|B|} \cdot {\mathbf f} \right] p_\parallel  - 2 |B| \left(\nabla_x \cdot \frac{{\mathbf f}}{|B|} \right) M_{2,1} \nonumber \\
[8pt]& & \hspace{0cm} + \nabla_x \cdot (K_{3,0} \, b) - 2 (b \cdot {\mathbb F}) K_{1,0} + 2 (b \cdot \nabla_x |B|) K_{1,1} = 0, \label{eq_p_par} 
\end{eqnarray}
and 
\begin{eqnarray}
& &  \hspace{-1cm} \frac{\partial p_\bot }{\partial t} + 
\nabla_x \cdot \left[ \left(u - \frac{b \times {\mathbb F}}{|B|} \right) \, p_\bot + \left( \, |B| \, (\nabla_x \times  \frac{b}{|B|}) - {\mathbf f} \, \right) \, |B| M_{0,2} + {\mathbf f} \, M_{2,1} \right] \nonumber \\
& & \hspace{0cm} + \left[ \, - (\nabla_x u) : (b \otimes b) +  \, \frac{{\mathbb F}}{|B|} \cdot {\mathbf f} 
+ \nabla_x \cdot u - \nabla_x \cdot \frac{b \times {\mathbb F}}{|B|} \right] p_\bot   \nonumber \\
& & \hspace{0cm} + \left(\nabla_x \cdot \frac{{\mathbf f}}{|B|} \right) |B|  M_{2,1}
+ |B| \nabla_x \cdot (K_{1,1} \, b) = 0. \label{eq_p_perp} 
\end{eqnarray}

\label{lem_np}
\end{lemma}

\noindent
{\bf Proof:} (i) The equation for the density is obtained by letting $m=q=0$ in (\ref{eq_Mmq}):
\begin{eqnarray}
&  & \hspace{-1cm} 
\frac{\partial}{\partial t} n  + \nabla_x \cdot \left[ 
\left(u - \frac{b \times {\mathbb F}}{|B|}\right) \, n + \left(|B|(\nabla_x \times \frac{b}{|B|}) - {\mathbf f} \right) \frac{p_\bot}{|B|} + \frac{{\mathbf f}}{|B|} p_\parallel \right] \, + \nonumber \\
&  & \hspace{9cm} 
+ \nabla_x \cdot (K_{1,0} \, b) = 0.  
\label{eq_n} 
\end{eqnarray}
But, using (\ref{nabla_Pi_pressions}), we get:
\begin{eqnarray*}
&  & \hspace{-1cm} 
\frac{b}{|B|} \times n {\mathbb F}   =  \frac{b}{|B|} \times  \nabla_x p_\bot +  \frac{{\mathbf f}}{|B|} (p_\parallel-p_\bot). 
\end{eqnarray*}
Therefore, we have: 
\begin{eqnarray*}
&  & \hspace{-1cm} 
- \frac{b \times n {\mathbb F}}{|B|} + \left(|B|(\nabla_x \times \frac{b}{|B|}) - {\mathbf f} \right) \frac{p_\bot}{|B|} + \frac{{\mathbf f}}{|B|} p_\parallel = \nabla_x \times \left( p_\bot \frac{b}{|B|} \right) ,
\end{eqnarray*}
and this term is canceled by the divergence operator in (\ref{eq_n}). With (\ref{Kflux=0}), eq. (\ref{Mass_cons}) follows.

Point (ii) follows from the application of the constraint (\ref{cont_Mmq}) for $m=0$ and $q=0$. Indeed, we find 
\begin{eqnarray*}
&  & \nabla_x \cdot (M_{1,0} \, b) =  0
 \, , 
\end{eqnarray*}
out of which (\ref{Int_G_cpar_0}) follows from the assumption on the boundary conditions. 

Point (iii) also follows from the inspection of the constraint (\ref{cont_Mmq}) but with $m=1$ and $q=0$. Indeed, the left-hand side of this equation is
\begin{eqnarray}
&  & \mbox{l.h.s.} = \nabla_x \cdot (p_\parallel \, b) -  b \cdot n {\mathbb F} + b \cdot \frac{\nabla_x {|B|}}{|B|} p_\bot 
 \, .  
\label{lhs}
\end{eqnarray}
But, thanks to (\ref{Pi_P}), (\ref{Pi_pressions}), we get 
\begin{eqnarray*}
&  & b \cdot n {\mathbb F} = b \cdot (\nabla_x \cdot {\mathbb P}) = b \cdot \nabla_x p_\parallel +   (p_\parallel - p_\bot) (\nabla_x \cdot b) 
 \, .  
\end{eqnarray*}
Therefore, (\ref{lhs}) is equal to
\begin{eqnarray*}
&  & \mbox{l.h.s.} = p_\bot (\nabla_x \cdot b) + b \cdot \frac{\nabla_x {|B|}}{|B|} p_\bot = \frac{p_\bot}{|B|} \nabla_x \cdot B = 0 
 \, ,  
\end{eqnarray*}
by the divergence free constraint on $B$. Therefore, the constraint (\ref{cont_Mmq}) for $m=1$ and $q=0$ is redundant with the definition of ${\mathbb F}$.

(iv) The equations for the pressures follow from the general moment equation (\ref{eq_Mmq}) and (\ref{nMKmq}). \endproof

Now, we can turn towards the main result of this section, namely the

\begin{lemma}
Let ${\mathcal G}$, ${\mathcal K}$ and $u$ satisfy eqs.~(\ref{gyro_cons_1}),~(\ref{gyro_cons_2}), (\ref{Kflux=0})
and~\eqref{vel_const_again} , then
\begin{equation}
u=u_{\parallel}\,b+u_{\bot},\label{vel_decomp}
\end{equation}
with $u_{\parallel}=u\cdot b$ and $u_{\bot}=b\times(u\times b)$ and $u_{\parallel}$ and
$u_{\bot}$ are solutions to~\eqref{transv_vel}, \eqref{eq_u_par}.
\label{lem_vel_expl}
\end{lemma}

\noindent
{\bf Proof:} 
In (\ref{vel_const_again}), we insert~\eqref{vel_decomp}, the decomposition of the velocity into its aligned and transverse parts.
By taking the vector product of (\ref{vel_const_again}) with $b$ and using (\ref{nabla_Pi_pressions}), we find that the transverse part of the velocity satisfies \eqref{transv_vel}. The first component is the classical $E \times B$ drift. The two other components is the expression of the diamagnetic drift when the parallel and transverse pressures are different. 

We now turn to the difficult part: the determination of $u_\parallel$. For it, we only have an implicit constraint, given by the projection of (\ref{vel_const_again}) onto $b$, i.e.
\begin{equation}
n (E \cdot b) - \Big( \, b \cdot \nabla_x p_\parallel + (p_\parallel - p_\bot) (\nabla_x \cdot b) \, \Big) = 0  \, .  
\label{par_vel}
\end{equation}

To show how this leads to a well-posed equation for $u_\parallel$, we take the time-derivative of (\ref{par_vel}) and use the continuity and pressure equations (\ref{Mass_cons}), (\ref{eq_p_par}), (\ref{eq_p_perp}) to eliminate the time derivatives of $n$, $p_\parallel$ and $p_\bot$. We first get from (\ref{par_vel}): 
\begin{eqnarray*}
&  & \frac{\partial n}{\partial t} \, (E \cdot b) + n \, \frac{\partial }{\partial t} (E \cdot b) - \left\{ \, b \cdot \nabla_x \frac{\partial p_\parallel}{\partial t} + \frac{\partial b}{\partial t}  \cdot \nabla_x p_\parallel + \right. \\
& & \hspace{4cm} \left.+ \frac{\partial }{\partial t}(p_\parallel - p_\bot) \, (\nabla_x \cdot b) + (p_\parallel - p_\bot) (\nabla_x \cdot \frac{\partial b}{\partial t} ) \, \right\} = 0  \, ,  
\end{eqnarray*}
which can be written 
\begin{eqnarray}
&  &  b \cdot \nabla_x \frac{\partial p_\parallel}{\partial t} - 
\frac{\partial n}{\partial t} (E \cdot b) + \frac{\partial }{\partial t}(p_\parallel - p_\bot) \, (\nabla_x \cdot b) =\nonumber \\
& & \hspace{4cm} n \, \frac{\partial }{\partial t} (E \cdot b) -\frac{\partial b}{\partial t}  \cdot \nabla_x p_\parallel-(p_\parallel - p_\bot) (\nabla_x \cdot \frac{\partial b}{\partial t} )  \, . 
\label{par_vel_dt_bis}
\end{eqnarray}
Now, from (\ref{Mass_cons}), we have 
$$ \frac{\partial n}{\partial t} = - \nabla_x \cdot (n u_\parallel b) - \nabla_x \cdot (n u_\bot)
\, .$$
Using the same methodology with eqs. (\ref{eq_p_par}), (\ref{eq_p_perp}), we find 
\begin{eqnarray*}
\frac{\partial p_\parallel}{\partial t} &=& - \nabla_x \cdot (p_\parallel u_\parallel b) - 2 \, p_\parallel  \, b \cdot \nabla_x u_\parallel + R_1,  \\
&=& - 3 \nabla_x \cdot ( p_\parallel  u_\parallel b)  + 2 u_\parallel \nabla_x \cdot (p_\parallel b) + R_1 \, ,  
 \\
\frac{\partial p_\bot}{\partial t} &=& - \nabla_x \cdot (p_\bot u_\parallel b) + p_\bot  b \cdot \nabla_x u_\parallel - p_\bot \nabla_x \cdot (u_\parallel b) + R_2, \\
&=&  - \nabla_x \cdot (p_\bot u_\parallel b)  - p_\bot u_\parallel (\nabla_x \cdot  b) + R_2
\, ,  
\end{eqnarray*}
where
\begin{eqnarray*}
& & R_1=-\nabla_x\cdot\left[p_{\parallel}\, \left(u_{\bot}-\frac{b \times {\mathbb F}}{|B|}\right) 
+ \left( \, |B| \, (\nabla_x \times  \frac{b}{|B|}\right) - {\mathbf f} \, ) \, M_{2,1} + \frac{{\mathbf f}}{|B|} \, M_{4,0}\right] \\
&&\hspace{2cm}+ 2 \left[ \, -(\nabla_x u_{\bot}) : (b \otimes b) +  \, \frac{{\mathbb F}}{|B|} \cdot {\mathbf f} \right] p_\parallel  + 2 |B| \left(\nabla_x \cdot \frac{{\mathbf f}}{|B|} \right) M_{2,1} \nonumber \\
& & \hspace{2cm} - \nabla_x \cdot (K_{3,0} \, b) + 2 (b \cdot {\mathbb F}) K_{1,0} - 2 (b \cdot \nabla_x |B|) K_{1,1}, 
\end{eqnarray*}
and
\begin{eqnarray*}
& & R_2=-\nabla_x \cdot \left[ \left(u_{\bot} - \frac{b \times {\mathbb F}}{|B|} \right) \, p_\bot + 
\left( \, |B| \, (\nabla_x \times  \frac{b}{|B|}) - {\mathbf f} \, \right) \, |B| M_{0,2} + {\mathbf f} \, M_{2,1} \right] \nonumber \\
& & \hspace{2cm} + \left[ (\nabla_x u_{\bot}) : (b \otimes b) -  \, \frac{{\mathbb F}}{|B|} \cdot {\mathbf f} 
- \nabla_x \cdot u_{\bot} + \nabla_x \cdot \frac{b \times {\mathbb F}}{|B|} \right] p_\bot   \nonumber \\
& & \hspace{2cm} - \left(\nabla_x \cdot \frac{{\mathbf f}}{|B|} \right) |B|  M_{2,1}
- |B| \nabla_x \cdot (K_{1,1} \, b). 
\end{eqnarray*}
Inserting these formulas into (\ref{par_vel_dt_bis}), we find~\eqref{eq_u_par} with $R_3$ given by 
\begin{eqnarray}
&& R_3=-(b\cdot\nabla_x)R_1-(E\cdot b)\nabla_x\cdot(n\,u_{\bot})-(\nabla_x\cdot b)\,(R_1-R_2)\nonumber\\
&&\hspace{3cm}+
n \, \frac{\partial }{\partial t} (E \cdot b) -\frac{\partial b}{\partial t}  \cdot \nabla_x p_\parallel-(p_\parallel - p_\bot) (\nabla_x \cdot \frac{\partial b}{\partial t} ).\label{R4}
\end{eqnarray}
This is an elliptic equation for $u_\parallel$ which is invertible provided boundary conditions for $u_\parallel$ are given at the ends of the magnetic field line. 

%%%%%%%%%%%%%%%%%%%%%%%%%%%%%%%%%%%%%%%%%%%%%%%%%%%%%%%%%%%%%%%
%%%%%%%%%%%%%%%%%%%%%%%%%%%%%%%%%%%%%%%%%%%%%%%%%%%%%%%%%%%%%%%
%%%%%%%%%%%%%%%%%%%%%%%%%%%%%%%%%%%%%%%%%%%%%%%%%%%%%%%%%%%%%%%
%%%%%%%%%%%%%%%%%%%%%%%%%%%%%%%%%%%%%%%%%%%%%%%%%%%%%%%%%%%%%%%
%%%%%%%%%%%%%%%%%%%%%%%%%%%%%%%%%%%%%%%%%%%%%%%%%%%%%%%%%%%%%%%

\setcounter{equation}{0}
\section{Appendix}\label{appen}

\subsection{Proof of Lemma~\ref{lem_Picgamma1}}
By (\ref{def_gyromoy}), (\ref{CV_c1})-(\ref{CV_c3}) we have in the frame $({\mathbf e}_1,{\mathbf e}_2,b)$:
\begin{eqnarray}
\Pi (c \, \gamma_1) &=& \frac1{2\pi} \int_{{\mathbb S}^1} \widetilde{(c \gamma_1)} (e, c_\parallel, \alpha) \,d\alpha, \nonumber \\
&=& \frac1{2\pi} \int_{{\mathbb S}^1} \left( \begin{array}{c} (2e - c_\parallel^2)^{1/2} \cos \alpha \\ (2e - c_\parallel^2)^{1/2} \sin \alpha \\ c_\parallel \end{array} \right)
\, 
\tilde \gamma_1 (\alpha) \,d\alpha.
\label{Picgamma1_1} 
\end{eqnarray}
But, thanks to (\ref{L=deriv_alpha}), we get:
\begin{eqnarray*}
\frac{1}{2\pi} \int_{{\mathbb S}^1}  \cos \alpha
\, 
\tilde \gamma_1 (\alpha) \,d\alpha &=& \frac1{2\pi} \int_{{\mathbb S}^1} \frac{d}{d\alpha} ( \sin \alpha)
\, 
\tilde \gamma_1 (\alpha) \,d\alpha,  \\
&=& - \frac{1}{2\pi} \int_{{\mathbb S}^1}  \sin \alpha
\, 
\frac{d \tilde \gamma_1}{d \alpha} (\alpha) \,d\alpha,  \\
&=& - \frac{1}{|B|} \frac{1}{2\pi} \int_{{\mathbb S}^1}  \sin \alpha
\, \, 
\widetilde {T g_0} (\alpha) \,d\alpha\, .  
\end{eqnarray*}
Similarly 
$$\frac{1}{2\pi} \int_{{\mathbb S}^1}  \sin \alpha
\, 
\tilde \gamma_1 (\alpha) \,d\alpha = \frac{1}{|B|} \frac{1}{2\pi} \int_{{\mathbb S}^1}  \cos \alpha
\, \, 
\widetilde {T g_0} (\alpha) \,d\alpha .$$
We introduce the new averaging operators, for an arbitrary function $h(c)$ and an arbitrary positive integer $m$:
\begin{eqnarray*}
\Pi_S^m h (e,c_\parallel) &=& \frac{1}{2\pi} \int_{{\mathbb S}^1}  \sin (m \alpha)
\, 
\tilde h (e,c_\parallel,\alpha) \,d\alpha ,  \\ 
\Pi_C^m h (e,c_\parallel) &=& \frac{1}{2\pi} \int_{{\mathbb S}^1}  \cos (m \alpha)
\, 
\tilde h (e,c_\parallel,\alpha) \,d\alpha ,
\end{eqnarray*}
which amounts to computing the $m$-th Fourier coefficients of $\tilde h$ with respect to $\alpha$. Then, the previous computation shows that 
$$\Pi (c \, \gamma_1) = \frac{1}{|B|} \, 
\left( \begin{array}{c} - (2e - c_\parallel^2)^{1/2} \Pi_S^1 T g_0 \\ (2e - c_\parallel^2)^{1/2} \Pi_C^1 T g_0 \\ 0 \end{array} \right).$$
The third line corresponds to the applications of the cancellation condition (\ref{cancel}). 

We now need to explicitly compute $Tg_0$ (so far, only $\Pi T g_0$ was computed). Using the specific form (\ref{ordre_0}) of $g_0$, we find:
\begin{eqnarray}
& & T g_0 = ({\mathbb F}_0 \cdot b) \frac{\partial G}{\partial c_\parallel}
+ c \cdot \left( \nabla_x + {\mathbb F}_0 \frac{\partial }{\partial e} \right) G
+ (c \otimes c) : \nabla_x b \,  \frac{\partial G}{\partial c_\parallel}. \label{exp_Tg0} 
\end{eqnarray}
Since $\Pi_{C,S}^1 (1) = 0$, we deduce that 
$$\Pi_{C,S}^1 T g_0 = 
\Pi_{C,S}^1 (c) \cdot \left( \nabla_x + {\mathbb F}_0 \frac{\partial }{\partial e} \right) G
+ \Pi_{C,S}^1 (c \otimes c) : \nabla_x b \,  \frac{\partial G}{\partial c_\parallel}, $$
and we are left with the task of computing $\Pi_{C,S}^1(c)$ and $\Pi_{C,S}^1 (c \otimes c) $. Using the same decomposition as for (\ref{Picgamma1_1}), we easily find: 
\begin{eqnarray*}
& & \Pi_S^1 (c) = \frac{1}{2} (2e-c_\parallel)^{1/2} \,  {\mathbf e}_2, \quad \Pi_C^1 (c) = \frac{1}{2} (2e-c_\parallel)^{1/2} \,  {\mathbf e}_1,\\
& &  \Pi_S^1 (c \otimes c) = \frac{1}{2} (2e-c_\parallel)^{1/2}  c_\parallel \, ({\mathbf e}_2 \otimes b + b \otimes {\mathbf e}_2), \\
& &  \Pi_C^1 (c \otimes c) = \frac{1}{2} (2e-c_\parallel)^{1/2}  c_\parallel \,  ({\mathbf e}_1 \otimes b + b \otimes {\mathbf e}_1) .
\end{eqnarray*}
Collecting these data, we deduce that:
\begin{eqnarray}
\Pi_S^1 T g_0  &=& \frac{1}{2}\, (2e -  c_\parallel^2)^{1/2} \, \left[ \Big(( \nabla_x + {\mathbb F}_0 \partial_e ) G \Big)_2  + c_\parallel \frac{\partial G}{\partial c_\parallel} \Big((\nabla_x b)_{23} + (\nabla_x b)_{32}\Big) \right], \label{PiS1Tg0} \\
\Pi_C^1 T g_0  &=& \frac{1}{2}\, (2e -  c_\parallel^2)^{1/2} \, \left[ \Big(( \nabla_x + {\mathbb F}_0 \partial_e ) G \Big)_1  + c_\parallel \frac{\partial G}{\partial c_\parallel} \Big((\nabla_x b)_{13} + (\nabla_x b)_{31}\Big) \right], \label{PiC1Tg0} 
\end{eqnarray}
and that
\begin{eqnarray}
\Pi (c \, \gamma_1) &=& \frac{1}{|B|} \, (e - \frac{1}{2} c_\parallel^2) \, \left\{
\, \left( \begin{array}{c} \displaystyle - (( \nabla_x + {\mathbb F}_0 \partial_e ) G )_2 \\ \displaystyle  (( \nabla_x + {\mathbb F}_0 \partial_e ) G )_1 \\ 0 \end{array} \right) \nonumber \right.\\
& & \left.\hspace{4cm} + c_\parallel \frac{\partial G}{\partial c_\parallel} 
\left( \begin{array}{c} \displaystyle - (\nabla_x b)_{23} -  (\nabla_x b)_{32} \\ \displaystyle (\nabla_x b)_{13} +  (\nabla_x b)_{31} \\ 0 \end{array} \right) \, \right\}
.
\label{Picgamma1_5} 
\end{eqnarray}
From the fact that $|b|=1$, we have $(\nabla_x b) b = 0$, out of which we deduce that $(\nabla_x b)_{i3} = 0$, for $i=1,2,3$. The first vector in  (\ref{Picgamma1_5}) can be easily identified with $b \times ( \nabla_x + {\mathbb F}_0 \partial_e ) G$ while the second one, which reduces to $(-(\nabla_x b)_{32}, (\nabla_x b)_{31}, 0)^T$ (the exponent $T$ denotes the transpose), is equal to 
$b \times ((b \cdot \nabla_x) b)$. Inserting these last remarks into (\ref{Picgamma1_5}) leads to (\ref{Picgamma1}) and ends the proof.

\subsection{Proof of Lemma~\ref{lem_Picocgamma1}}
Using the same method as in the previous lemma, we show that in the basis $({\mathbf e}_1,{\mathbf e}_2,b)$, the matrix $\Pi ((c \otimes c) \, \gamma_1)$ has the expression:
$$\Pi ((c \otimes c) \, \gamma_1) = \frac{1}{|4B|} \, 
\left( \begin{array}{ccc} - (2e - c_\parallel^2) \Pi_S^2 T g_0 & \times & \times
\\ 
(2e - c_\parallel^2) \Pi_C^2 T g_0 & (2e - c_\parallel^2) \Pi_S^2 T g_0 & \times \\ - 4(2e - c_\parallel^2)^{1/2} c_\parallel  \Pi_S^1 T g_0 & 4(2e - c_\parallel^2)^{1/2} c_\parallel  \Pi_C^1 T g_0 & 0 
\end{array} \right),$$
where the symbol $\times$ indicates that the matrix is symmetric. So, again, we are left with the computation of $\Pi_{C,S}^2 T g_0$. In view of (\ref{exp_Tg0}) and the fact that, obviously, $\Pi_{C,S}^2 (1) = \Pi_{C,S}^2 (c) = 0$, we need to compute $\Pi_{C,S}^2 (c \otimes c)$. The same method as previously applies and leads to
\begin{eqnarray*}
& & \Pi_{S}^2 (c \otimes c) = \frac{2e-c_\parallel^2}{4} \, \, ({\mathbf e}_1 \otimes {\mathbf e}_2 + {\mathbf e}_2 \otimes {\mathbf e}_1), \\
& & \Pi_{C}^2 (c \otimes c) = \frac{2e-c_\parallel^2}{4} \, \,  ({\mathbf e}_1 \otimes {\mathbf e}_1 - {\mathbf e}_2 \otimes {\mathbf e}_2). 
\end{eqnarray*}
We deduce that 
\begin{eqnarray*}
& & \Pi_{S}^2 T g_0 = \Pi_{S}^2 (c \otimes c) : \nabla_x b \, \, \frac{\partial G}{\partial c_\parallel} =  \frac{2e-c_\parallel^2}{4} \, \frac{\partial G}{\partial c_\parallel} \,\, ( (\nabla_x b)_{12} + (\nabla_x b)_{21}), \\
& & \Pi_{C}^2 T g_0 = \Pi_{C}^2 (c \otimes c) : \nabla_x b \, \, \frac{\partial G}{\partial c_\parallel} =  \frac{2e-c_\parallel^2}{4} \, \frac{\partial G}{\partial c_\parallel} \,\, ( (\nabla_x b)_{11} - (\nabla_x b)_{22}).  
\end{eqnarray*}

With (\ref{PiS1Tg0}) and (\ref{PiC1Tg0}), we deduce that $\Pi ((c \otimes c) \, \gamma_1)$ has the expression in the basis $({\mathbf e}_1,{\mathbf e}_2,b)$:
\begin{eqnarray}
\Pi ((c \otimes c) \, \gamma_1) &=& \frac{(2e - c_\parallel^2)^2}{16|B|} \, \frac{\partial G}{\partial c_\parallel} \, 
\left( \begin{array}{ccc} -(\nabla_x b)_{12} - (\nabla_x b)_{21} & \times & \times \\
(\nabla_x b)_{11} - (\nabla_x b)_{22} & (\nabla_x b)_{12} + (\nabla_x b)_{21} & \times \\ 
0 & 0 & 0 
\end{array} \right) \nonumber \\
& + &  
\frac{(2e - c_\parallel^2) c_\parallel}{2|B|} \, \, 
\left( \begin{array}{ccc} 0 & \times & \times \\
0 & 0 & \times \\ 
- (( \nabla_x + {\mathbb F}_0 \partial_e ) G )_2 &  (( \nabla_x + {\mathbb F}_0 \partial_e ) G )_1 & 0 
\end{array} \right) \nonumber \\
& + &  
\frac{(2e - c_\parallel^2) c_\parallel^2}{2|B|} \, \frac{\partial G}{\partial c_\parallel}\, 
\left( \begin{array}{ccc} 0 & \times & \times \\
0 & 0 & \times \\ 
- ( \nabla_x b)_{32} &  ( \nabla_x b)_{31} & 0 
\end{array} \right).
\label{Picocgamma1_2} 
\end{eqnarray}
Now, we need to evaluate $(\nabla_x b):\Pi ((c \otimes c) \, \gamma_1)$. It is an easy matter to see that the contracted product of $(\nabla_x b)$ with the first matrix of (\ref{Picocgamma1_2}) is identically zero, as well as with the third one. The contracted product of  $(\nabla_x b)$ and the second matrix of (\ref{Picocgamma1_2}) involves the expression 
\begin{eqnarray}
& &  - (\nabla_x b)_{31} \, \Big((\nabla_x + {\mathbb F}_0 \partial_e )G\Big)_2 + (\nabla_x b)_{32} \, \Big((\nabla_x + {\mathbb F}_0 \partial_e )G\Big)_1 = \nonumber \\
& & \hspace{5cm} =  b \cdot \left\{ 
\Big((\nabla_x + {\mathbb F}_0 \partial_e )G\Big) \times (b \cdot \nabla_x)b
\right\},  \nonumber \\
& & \hspace{5cm} =  - \Big((\nabla_x + {\mathbb F}_0 \partial_e )G\Big) \cdot \Big( 
b \times (b \cdot \nabla_x)b\Big).
\label{Picocgamma1_3} 
\end{eqnarray}
Collecting (\ref{Picocgamma1_2}) and (\ref{Picocgamma1_3}) leads to the result (\ref{Picocgamma1}) and ends the proof of the lemma.

\vspace{0.2cm}

{\bf Acknowledgments.}~The authors wish to express their gratitude
to G.~Falc\-hetto, X.~Garbet and M.~Ottaviani from the CEA-Cadarache and F.
Deluzet from the Institut de Math\'ema\-tiques de Toulouse,  for
fruitful discussions and encouragements. This work has been partially
supported by the Marie Curie Actions of the European Commission in the
frame of the DEASE project (MEST-CT-2005-021122), by the CNRS and the
Association Euratom-CEA in the framework of the contract 'Gyrostab'
and by the CEA-Saclay in the framework of the contract 'Astre' \# SAV
34160. This work was performed while the second author was an INRIA-Post-Doc 
at the Institut de Math\'ematiques de Toulouse.  

%%%%%%%%%%%%%%%%%%%%%%%%%%%%%%%%%%%%%%%%%%%%%%%%%%%%%%%%%%%%
%%%%%%%%%%%%%%%%%%%%%% 
%%%%%%%%%%%%%%%%%%%%%%%%%%%%%%%%%%%%%%%%%%%%%%%%%%%%%%%%%%%%

\end{document}